\documentclass{article}
\makeatletter
\def\ps@pprintTitle{%
 \let\@oddhead\@empty
 \let\@evenhead\@empty
 \def\@oddfoot{}%
 \let\@evenfoot\@oddfoot}
\makeatother
\usepackage{makecell}
\usepackage{float}
\usepackage{fancyhdr}
\newcommand{\dd}{\ensuremath{\mathrm{d}}}

\newcommand*\colvec[3][]{
	\begin{pmatrix}\ifx\relax#1\relax\else#1\\\fi#2\\#3\end{pmatrix}
}
\usepackage{amsmath}            
\usepackage{comment}

\usepackage{epstopdf}           
\usepackage{flushend}           

\usepackage{amssymb}
\usepackage{scalerel}
 \usepackage{multirow}
 \usepackage{graphicx}

\usepackage[figuresright]{rotating}

\usepackage[round]{natbib}
\usepackage{subfigure}
\usepackage{authblk}

\providecommand{\keywords}[1]
{
  \small	
  \textbf{\textit{Keywords---}} #1
}

\title{Multiscaling and rough volatility: an empirical investigation}

\date{}

\author[1]{Giuseppe Brandi\thanks{Corresponding author:\\ 
		Email address: giuseppe.brandi@kcl.ac.uk (Giuseppe Brandi)}}

\author[1,2,3]{T. Di Matteo}

\affil[1]{Department of Mathematics, King's College London, WC2R 2LS London, UK}
\affil[2]{Complexity Science Hub Vienna, 1080 Vienna, Austria}
\affil[3]{Centro Ricerche Enrico Fermi, Via Panisperna 89 A, 00184 Rome, Italy}

\begin{document}

  \maketitle

\begin{abstract}
Pricing derivatives goes back to the acclaimed Black and Scholes model. However, such a modeling approach is known not to be able to reproduce some of the financial stylized facts, including the dynamics of volatility. In the mathematical finance community, it has therefore emerged a new paradigm, named rough volatility modeling, that represents the volatility dynamics of financial assets as a fractional Brownian motion with Hurst exponent very small, which indeed produces rough paths. At the same time, prices' time series have been shown to be multiscaling, characterized by different Hurst scaling exponents. This paper assesses the interplay, if present, between price multiscaling and volatility roughness, defined as the (low) Hurst exponent of the volatility process. In particular, we perform extensive simulation experiments by using one of the leading rough volatility models present in the literature, the rough Bergomi model. A real data analysis is also carried out in order to test if the rough volatility model reproduces the same relationship. We find that the model is able to reproduce multiscaling features of the prices' time series when a low value of the Hurst exponent is used but it fails to reproduce what the real data say. Indeed, we find that the dependency between prices' multiscaling and the Hurst exponent of the volatility process is diametrically opposite to what we find in real data, namely a negative interplay between the two.

\end{abstract}






\section{Introduction}

The history of derivatives pricing goes back to the famous Black and Scholes model \citep{black1973,merton1973theory}. In the literature, several models added robustness to this original model by trying to adapt it more to reality. In particular, some models have introduced the direct modeling of the volatility dynamics of the diffusive price process \citep{heston1993closed}. This modeling approach has the advantage to incorporate in the pricing procedure the dynamics of the volatility, avoiding the strong assumption of constant volatility. Moreover, in order to accommodate the stylized facts for which the volatility and price dynamics are empirically negatively dependent, a correlation parameter has been introduced between the Brownian motion that drives the two dynamics \citep{heston1993closed}. Still, these features are not able to depict some aspects of the empirical data, i.e. the implied volatility surface \citep{roughvola,rBergomi}. For this reason, \citep{roughvola} has introduced the concept of rough volatility. In this setting, the volatility dynamics is depicted as a fractional process (a fractional Brownian motion), with a very small Hurst exponent (the long-memory parameter). This is supported by the empirical analysis of Realized Variance (RV) measures\footnote{Realized variance is the sum of squared returns over a specific time window for a specific time frequency. For example, the RV can be the sum of squared intra-day returns at 10 minutes frequency, which is an estimate of price variation over the day.} estimated by using high frequency data. In fact, several papers have found that realized variance has a Hurst parameter very small, i.e. $H\sim0.1$, i.e. volatility is rough. This new formulation is able to reproduce implied volatility surface dynamics more accurately. \par
However, in all these models the stochastic process that drives the price dynamics is a standard Brownian motion, i.e. a process with Hurst parameter equal to $0.5$. Indeed, log-prices have been empirically shown to deviate from the Brownian motion paradigm in two main aspects. First, the long-memory parameter, the Hurst exponent, is not (statistically) equal to $0.5$, and second, in contrast with the Brownian motion, different statistical moments yield different Hurst exponents, i.e. financial time series are multiscaling.
Multiscaling is by now identified as stylized facts in financial time series. The study of scaling and multiscaling have been a prominent topic in quantitative finance literature which devoted most of the attention to financial time series in order to understand the source of multiscaling from an empirical and theoretical point of view \citep{mandelbrot1963,mandelbrot1967variation,dacorogna_book,mantegna_stanley,scaling_review_tiziana,calvet1,lux1,lux_marchesi,tiziana_dacorogna2,buonocore2020}.  
The estimation of multiscaling properties is challenging and it requires robust statistical procedures \citep{brandi2021}.  \par
Both multiscaling and rough volatility have been understood to originate from one or more phenomena related to trading dynamics but unlike the analysis of the prices-volatility dependence, which has been shown to be strongly negative, to our knowledge, no dependency analysis of their scaling properties' has been produced so far. In particular, an important point is to investigate if rough volatility models are able to produce the multiscaling features empirically found in prices time series and to study their interplay with volatility roughness  (defined as the Hurst exponent of the volatility process). This might have strong implications for modeling prices behaviors and risk forecasting since by calibrating a wrong interplay, the degree of multiscaling in the price process would be under- or over-estimated. \par
In this paper, we fill this gap by studying the dependency between rough volatility and prices' multiscaling by using one of the benchmark rough volatility models, namely the rough Bergomi model \citep{rBergomi}, and check if it is able to reproduce multiscaling and the same scaling structure of the real data. To this end, we first compute the Hurst exponent of the volatility process and the multiscaling measure of the price time series by using the methodology proposed in \citep{brandi2021} and then compute a set of correlation coefficients between the two measures. To also check if results are affected by outliers, we also use an outlier-robust correlation estimation methodology \citep{wilcox2004inferences,wilcox2011introduction,pernet2013robust,wilcox2018improved}. The paper is structured as follows. Section \ref{fbm} and Section \ref{rvola} review some concepts of fractional Brownian motion and rough volatility. Section \ref{sec_ms} reports the statistical procedures used to estimate multiscaling and the correlation analysis. Section \ref{sec_ea} shows results of a simulation experiment by using the rough Bergomi model while Section \ref{sec_se} those of the dependency analysis between prices' multiscaling and volatility roughness for real data. Section \ref{sec_c} concludes.

\section{Fractional Brownian motion}\label{fbm}

Historically, the Black-Scholes (BS) model for option pricing \citep{black1973,merton1973theory}, has been (and still is) the cornerstone in quantitative finance. By means of the Geometric Brownian motion, the authors were able to provide an equation that can be used to compute the price of vanilla options. However, some researchers questioned the BS model's assumptions \citep{mandelbrot1967variation}. In particular, one of these assumptions is the adoption of a Brownian motion for the price fluctuations, which implies no memory (Markovian property). A possible solution to this inconsistency with respect to the real data was identified in replacing the Brownian motion with a fractional Brownian motion \citep{mandelbrot1967variation,mandelbrot1968noah}.
A fractional Brownian motion is a stochastic process characterized by the following three properties \citep{taqqu2013benoit}:
\begin{itemize}
\item[(1)] the process is Gaussian with zero mean;
\item[(2)] it has stationary increments;
\item[(3)] it is self-similar with index $H$, $0<H<1$.
\end{itemize}
Fractional Brownian motion reduces to Brownian motion when $H=1/2$, but in contrast to Brownian motion, it has dependent increments when $H \neq1/2$, i.e. it is a non-Markovian process.\footnote{Pricing derivatives under non-Markovianity is very challenging and Monte-Carlo procedures are usually employed.}
In order to compute the Hurst exponent from sample data, in this paper we use the method of \citep{brandi2021} that is based on Generalized Hurst Exponent method (GHE), see \citep{scaling_review_tiziana,kantelhardt,tiziana_dacorogna,tiziana_dacorogna2, buonocore,buonocore2,antoniades2020use}. This methodology relies on the measurement of the direct scaling of the $q$th-order moments of the distribution of the increments (described in Section \ref{sec_ms}). The GHE methodology returns the scaling exponent $H_q$. The most relevant (and used) values of $q$ to assess the scaling properties of a time series are $q=1$ and $q=2$. The first one depicts the scaling of the absolute values of the increments and is closely related to the Hurst exponent originally proposed by \citep{hurst1956methods} while the second is associated to the scaling of the autocorrelation function of the process \citep{scaling_review_tiziana}. In the remainder of the paper, when not specified differently, we refer to $H$ as the Hurst exponent computed for $q=1$.

\section{Rough volatility}\label{rvola}

Building on the work of \citep{roughvola} on the statistical analysis of realized variance, rough volatility became a new paradigm in quantitative finance. It has been shown that realized variance (a proxy for rough volatility) is characterized by a process rougher that Brownian motion, i.e. $H<\frac{1}{2}$. This empirical observation lead to the construction of stochastic models with strong anti-persistent volatility dynamics, the so called rough volatility models \citep{roughvola,roughvola2,roughvola3,roughvola4}.
One of the most leading models in this category is the rough Bergomi model (hereafter rBergomi), see \citep{rBergomi,rBergomi_calibration}. In the rBergomi model, the dynamics for the asset price process $S_t$ and the instantaneous variance process $v_t$ are given by
	\begin{align}\label{eq_rb}
	\frac{\dd S_t}{S_t} &= \sqrt{v_t} \dd \left(\lambda W_t + \sqrt{1-\lambda^2}W^{\bot}_t \right) \\
	v_t &= \xi_0 \exp\left(\eta W_t^H  - \frac{1}{2} \eta^2 t^{2H}\right), \quad t \in [0,T].
		\end{align}
	Here $W_t$ and  $W_t^{\bot}$ are two independent Brownian motions, $T$ is the final time step, $\lambda \in [-1,1] $ is the correlation parameter between the price and volatility dynamics, $\eta > 0$ denotes the volatility of the volatility process, and $\xi_0(t)$ is the initial forward variance curve and $H$ is the Hurst exponent. Moreover, $W^H$ is a fractional Brownian motion given by
	\begin{equation}
		W_t^H = \sqrt{2H}\int_0^t (t-s)^{H-\frac{1}{2}} \dd W_s, \quad t \in [0,T],
	\end{equation}
	with the Hurst parameter being $H \in (0,1)$. Generally, rough volatility models are calibrated for $H$ to be very small, i.e. $H\sim0.1$.  Due to the lack of Markovianity, conventional analytical pricing methods cannot be employed and Monte Carlo pricing methods based on simulated paths are used instead \citep{mccrickerd2018turbocharging}.

\section{Multiscaling}\label{sec_ms}
 In Section \ref{fbm}, we have recalled the fractional Brownian motion and the estimation of the Hurst exponent. However, financial time series have been shown to be not only scaling but also multiscaling \citep{scaling_review_tiziana}. To detect multiscaling, it is necessary to study the non-linearity of the scaling exponents of the $q$-order moments of the absolute value of the process' increments \citep{mandelbrot1,calvet3,scaling_review_tiziana}. 
In particular, for a process $X(t)$ with stationary increments (at time aggregation $\tau$) $r_{\tau}(t)$, i.e. $r_{\tau}(t)$=$X(t+\tau)-X(t)$, the GHE methodology considers a function of increments \citep{scaling_review_tiziana} of the form 
 \begin{equation}\label{mult_def}
\Xi(\tau,q)= \mathbb{E}\left[|r_{\tau}(t)|^q\right]\sim K_q\tau^{qH_q},
 \end{equation}
 where $q=\{q_1,q_2,\dots,q_M\}$ is the set of evaluated moments, $\tau=\{\tau_1,\tau_2,\dots,\tau_N\}$ is the set of time aggregations used to compute the log-returns, $N$ and $M$ are the maximum numbers of moments and time aggregation' specifications, i.e. $q_1=q_{min}$, $q_M=q_{max}$, $\tau_1=\tau_{min}$ and $\tau_N=\tau_{max}$, $K_q$ is the $q$-moment for $\tau=1$, and $H_q$ is the so called generalized Hurst exponent which is a function of $q$.
Recently \citep{brandi2021} proposed to compute the value of $K_q$ by evaluating $\Xi(1,q)$ rather than estimating it via regression in order to remove any possible bias introduced in the estimation. By normalizing the structure function $\Xi(\tau,q)$ as 
 \begin{equation}\label{new_model}
\widetilde{\Xi}(\tau,q)= \frac{\Xi(\tau,q)}{K_q},
 \end{equation}
 Equation \ref{new_model} eliminates the possible bias introduced by the estimation of $K_q$ via regression. Further, the q-order normalized moment is defined as
  \begin{equation}\label{new_model3}
 \dddot{\Xi}(\tau,q)=\widetilde{\Xi}(\tau,q)^{\frac{1}{q}}
 \end{equation}
 from which follows that Equation \ref{mult_def} becomes
  \begin{equation}\label{mult_def3}
 \dddot{\Xi}(\tau,q)\sim \tau^{H_q}.
 \end{equation}
Within this new formulation, the $q$ regressions have a $0$ intercept and the multiscaling is present only if the regression coefficients $H_q$ differ for distinct values of $q$. To assess multiscaling, it is then possible to analyze the equation of the form
\begin{equation}\label{mult_proxy3}
H_q=A+Bq.
\end{equation}
where $A$ is the linear scaling index while $B$ is the multiscaling proxy. In this mathematical setting, as for different multifractal models in finance \citep{bacry1,calvet1,calvet4,sornette2018multifractal}, we implicitly assume a quadratic function of $qH_q$. Eliminating the multiplication by $q$ from both sides of Equation \ref{mult_proxy3}, we reduce the possibility of spurious results in case $q$ is a dominant factor in the multiplication. By estimating $B$, and testing its statistical significance, we are statistically able to identify multiscaling time series. In the following, we will refer to the scaling measures of the volatility process with the superscript $(v)$ and with the superscript $(P)$ for the prices, e.g $H^{(v)}$, $B^{(v)}$, $H^{(P)}$ and $B^{(P)}$.

\subsection{The choice of $\tau_{max}$}\label{sec_taumax}
As reported in \citep{brandi2021}, the choice of the maximum aggregation time is pivotal for the correct estimation of the scaling exponents and by consequence, the multiscaling properties. This pivotal choice is mainly due to the fact that in real data, even for very persistent time series, there is an aggregation cutoff from which the financial time series behave as uncorrelated. If we choose arbitrarily the maximum value of the aggregation time, we could mix long-range correlation with an uncorrelated state, producing an erroneous estimation of the scaling exponents. To this extent, several methodologies have been proposed in the literature \citep{sornette2018multifractal}. In this paper, we use the Autocorrelation Segmented Regression proposed in \citep{brandi2021}. The idea of this approach is to perform a segmented regression on the autocorrelation (or the autocovariance) function computed on the absolute returns and take $\tau_{max}=\tau^{*}$ as the splitting point between the long-range dependence state and the random state which minimizes the sum of squared residuals. By using the Autocorrelation Segmented Regression (ACSR), we can write the autocorrelation function of the absolute returns $r_{\tau}(t)$ for lag $\tau$, $\phi_{\tau} (|r_{\tau}(t)|)$ as:
\begin{equation}
\phi_{\tau} (|r_{\tau}(t)|) = \begin{cases} 
\alpha+\tau^{\beta}, & \text{if $\tau<\tau_{*}$} \\
\alpha+\tau_{*}^{\beta}, & \text{if $\tau \geq \tau_{*}$} 
\end{cases}
\end{equation}
where $\alpha$ is the intercept of the regression and that can be fixed to be equal to $\phi_{1}$, $\beta$ is a memory exponent for the autocorrelation function, $\tau$ is the lag at which the autocorrelation is computed, and $\widehat{\tau}^{*}$ is the estimated value of aggregation which split the autocorrelation function between the correlated and random states. This estimated parameter will be used as $\tau_{max}$ in the GHE estimation procedure.  

\subsection{Analysis of dependence}

In this paper, we are interested in analyzing the dependency between the scaling and multiscaling measures of the volatility time series and the scaling and multiscaling measures of the price process. Among the different measures that are available in the statistical literature, we use the Pearson and Spearman correlation coefficients. The Pearson correlation coefficient for any two random variables $X$ and $Y$ is defined as

\begin{equation}\label{corr}
  \rho=\rho(X,Y)=\frac{Cov(X,Y)}{\sigma_{X}\sigma_{Y}},
\end{equation}

where Cov($X$,$Y$) is the covariance between $X$ and $Y$ and $\sigma_{X}$ and $\sigma_{X}$ are the standard deviation of $X$ and $Y$, respectively. Contrary to the Pearson correlation, Spearman correlation captures the monotonic dependency (linear or nonlinear) between the two variables analyzed. Indeed, the Spearman correlation corresponds to the Pearson correlation between the rank values of the random variables. Let us define as $R(X)$ the ranks of $X$, the Spearman correlation is defined as:

\begin{equation}\label{corr2}
  \rho_S=  \rho(R(X),R(Y))=\frac{Cov(R(X),R(Y))}{\sigma_{R(X)}\sigma_{R(Y)}},
\end{equation}

where the quantities are defined as for the Pearson correlation.

\section{The interplay between multiscaling and rough volatility: synthetic data}\label{sec_ea}

\begin{figure}[h]
		\begin{center}	
			
			\includegraphics[width=0.9\textwidth,height=0.4\textheight]{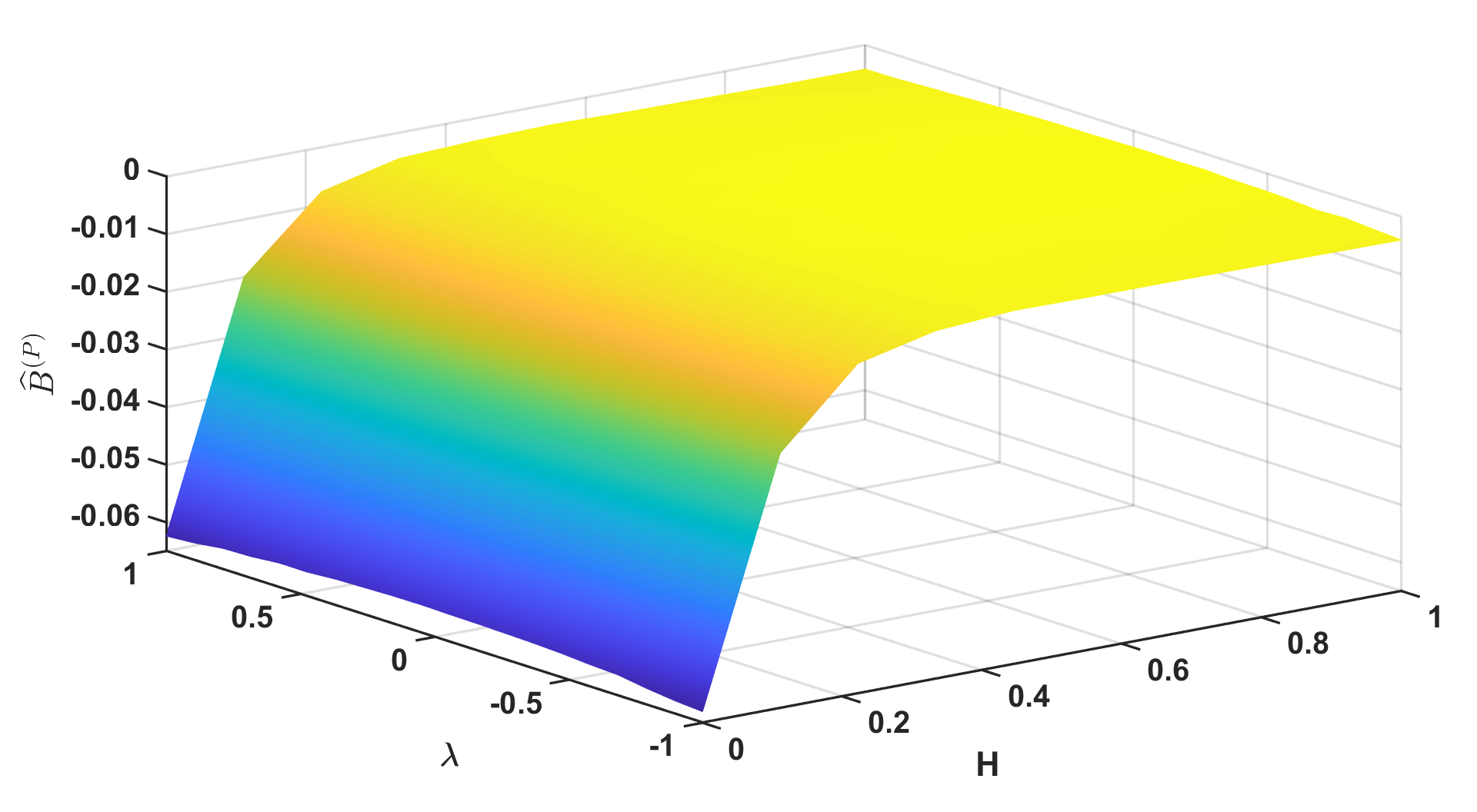} 
		\end{center}
		\caption{Multiscaling proxy $\widehat{B}^{(P)}$ as a function of $H$ and $\lambda$ in the rBergomi model. The result is averaged over the $100$ datasets and the plot is smoothed via interpolation for better representation.}
		\label{fig5c}
	\end{figure}

In this section, we simulate the rough Bergomi (rBergomi) model \citep{rBergomi,rBergomi_calibration} and check if it is able to produce multiscaling prices. If it is the case, we want to understand what type of dependency structure there is between the model's parameters and the simulated prices multiscaling.\footnote{We report in \ref{add_res_sin} also the analysis done with respect to other scaling measures.}
For the simulations, we use the parameters used in \citep{rBergomi}, i.e. $\eta=1.9$, $\xi_0=0.1$ and varying values of the Hurst exponent $H$ and correlation parameter $\lambda$ in Equation \ref{eq_rb}. We are interested in analyzing the model's potentiality to generate a dependency structure between the scaling measures of the two processes by changing the two parameters, $H$ and $\lambda$. In particular, we set the correlation parameter $\lambda$ to vary between $-1$ and $+1$ and the Hurst exponent $H$ to vary between $0.01$ and $0.99$. To mimic the real data structure presented in \ref{sec_data}, and take into account any possible finite sample effect, for each combination of $H$ and $\lambda$, we simulate $100$ sets of time series (volatility and prices), each of which has been taken with the same lengths as the ones of the original dataset, i.e. $T=5000$ time steps. We then compute $H^{(v)}$ and $B^{(P)}$ of the two simulated processes and analyze their dependence. Figure \ref{fig5c} shows the impact of the Hurst parameter $H$ and correlation parameter $\lambda$ on the estimated multiscaling proxy $B$, $\widehat{B}^{(P)}$.

\begin{figure}[H]
		\begin{center}	
			
			\includegraphics[width=0.9\textwidth,height=0.34\textheight]{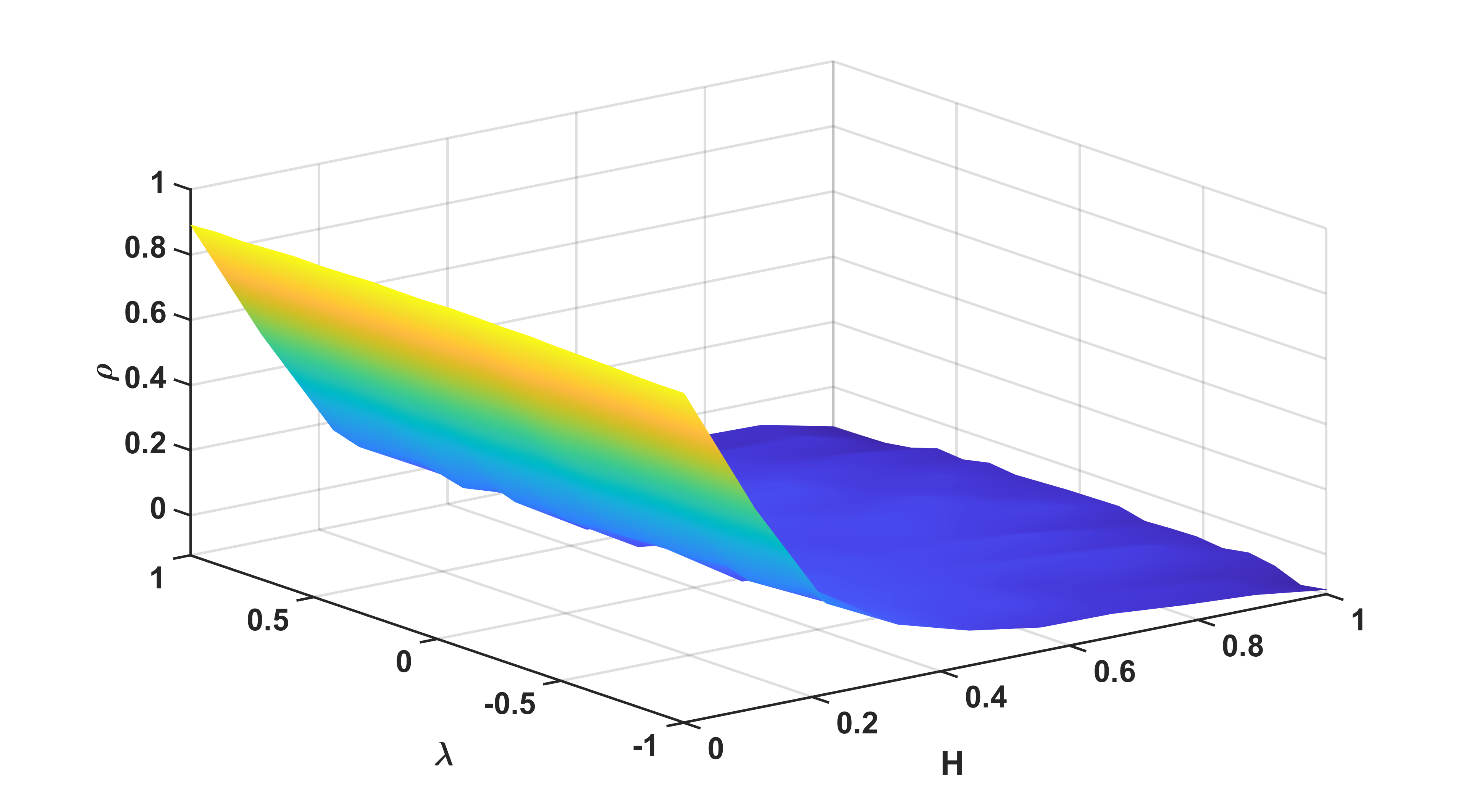} 
		\end{center}
		\caption{Pearson correlation between Multiscaling proxy $\widehat{B}^{(P)}$ and $\widehat{H}^{(v)}$ as function of $H$ and $\lambda$ in the rBergomi model. The result is averaged over the $100$ datasets and the plot is smoothed via interpolation for better representation.}
		\label{fig6a}
	\end{figure}

Figure \ref{fig5c} shows that the rBergomi model is indeed able to produce multiscaling prices for small values of $H$ irrespective of the value of $\lambda$. Indeed, the effect of $\lambda$ is very small and almost negligible. As it is possible to notice, the dependency relationship is stronger for small values of $H$ while it becomes negligible for high values of $H$. To numerically quantify this finding, we compute the Pearson and Spearman correlations between the volatility roughness $\widehat{H}^{(v)}$ and the price multiscaling $\widehat{B}^{(P)}$ for the $100$ sets of time series. To better understand the local behavior of the dependency, we have partitioned the entire set of $H$ values in $10$ subsets with width $0.1$, i.e. $\{(0,0.01,\dots,0.10),(0.10,0.11,\dots,0.20),\dots,(0.90,0.91,\dots,1)\}$ and we have computed the correlation coefficients between the estimated $H^{(v)}$ on each subset and the corresponding estimated values of the prices' multiscaling proxy $B^{(P)}$.\footnote{We repeated the same exercise with respect to $\lambda$, but we find an erratic behavior as $\lambda$ does not play a significant role. For this reason, we do not report the plot.} Results of the averaged Pearson correlations computed over the $100$ simulated sets of times series are shown in Figures \ref{fig6a}.\footnote{We report the same analysis with respect to the Spearman correlation in Figure \ref{fig_apx2} of \ref{add_res_sin}. The results are qualitatively equivalent to the Pearson correlation.}

Figure \ref{fig6a} confirms what was deduced from Figure \ref{fig5c}. The correlation between volatility roughness and prices multiscaling diminishes as the Hurst exponent $H$ of the volatility process increases. Furthermore, the correlation becomes negligible already for $H$ near $0.3$.\footnote{The correlation for values equal or higher than $0.3$ is not statistically significant at $5\%$.} From this exploratory analysis we might conclude that in order to retrieve both multiscaling and interplay between $\widehat{H}^{(v)}$ and $\widehat{B}^{(P)}$ , we would need to use very low values of the Hurst exponent in Equation \ref{eq_rb}, while the value of the parameter $\lambda$ does not have a strong impact on the interplay.


\section{The interplay between multiscaling and rough volatility: Real data}\label{sec_se}
In the previous Section, we have shown that the rBergomi model is able to generate multiscaling prices when small values of $H$ are used in Equation \ref{eq_rb}, irrespective of the model's correlation parameter $\lambda$. We also found an overall nonlinear relationship between the level of multiscaling and the Hurst exponent $H$.   
In this section, we repeat a similar exercise as the one done in Section \ref{sec_ea} and compute the correlation coefficients between the volatility roughness and price multiscaling of real data. This exercise is allowing us to understand if rough volatility models are in line with the empirical observations. In particular, by using data from the Oxford volatility library \citep{library},\footnote{\ref{sec_data} reports the description of the dataset.} we first compute the Hurst exponent $H$ on the realized variance (10 min frequency) time series, i.e. $\widehat{H}^{(v)}$ and the multiscaling proxy $B$ on the prices time series, i.e, $\widehat{B}^{(P)}$. We then produce a set of correlation measures to quantify their interplay.\footnote{We also produced results for other scaling measures and for different rough volatility proxies. Results are reported in \ref{add_res_real}.}

\subsection{Results}\label{res_procedure}
In this section, we report the procedure used on real data to compute the scaling exponents and the multiscaling proxy, defined as follows:

\begin{enumerate}
	\item We first compute $\tau^{*}$ with the Autocorrelation Segmented Regression method introduced in Section \ref{sec_taumax} by using the absolute value of the open to close log-returns;\footnote{Using close to close log-returns the results remain qualitatively unchanged.}
	\item We then perform the log-log regression of Equation \ref{mult_def3} for each index with $\tau_{max}=\widehat{\tau}^{*}$, that is the estimated $\tau^*$.\footnote{We use $q_{min}=0.05$ and $q_{max}=1$ as prescribed in \citep{buonocore2020,brandi2021}.} 
	\item We finally compute the multiscaling proxy $\widehat{B}$ for each index by using Equation \ref{mult_proxy3} and test for its statistical significance.
\end{enumerate}    

Results of this procedure for the rough volatility measure (Realized Variance 10 minutes frequency) are reported in Table \ref{res_scaling}. A set of preliminary conclusions can be drawn from these results. First of all, it can be appreciated that there is heterogeneity in terms of optimal aggregation time even if many indices fall in the range between 1 and 3 trading years, with an average of 2 trading years. The second piece of evidence that can be extracted from Table \ref{res_scaling} is that the volatility is indeed rough with a Hurst exponent ($\widehat{H}^{(v)}$ in the table) between $\sim0.08$ and $\sim0.15$ and that rough volatility presents very low (negligible) multiscaling values, as reported in other research papers \citep{roughvola,roughvola4}. In contrast to the realized variance time series, the prices time series present a much stronger multiscaling feature ($\widehat{B}^{(P)}$ in the table) across all markets, confirming what was found in a set of recent papers \citep{buonocore2020,brandi2021}. Finally, it is possible to notice that apart from some cases, the Hurst exponent for prices is different from the $0.5$ benchmark.\footnote{It is important to highlight the fact that being $H$ an exponent, even small deviations from $0.5$ are influential.}

\begin{table}[H]
\centering
\resizebox{0.8\textwidth}{0.35\textheight}{%
\begin{tabular}{|l|l|l|l|l|l|} 
\hline
\multicolumn{1}{|c|}{\textbf{Index}} & \multicolumn{1}{c|}{\textbf{}}& \multicolumn{2}{c|}{\textbf{Prices}} & \multicolumn{2}{c|}{\textbf{Volatility}}  \\ 
\hline
                  & $\tau_*$    & $\widehat{H}^{(P)}$ & $\widehat{B}^{(P)}$                        & $\widehat{H}^{(v)}$ & $\widehat{B}^{(v)}$                                    \\ 
\hline
\textbf{AEX}      & 516         & 0.523 & -0.027                       & 0.130 & -0.008                                   \\ 
\hline
\textbf{AORD}     & 507         & 0.510 & -0.004                       & 0.061 & -0.007                                   \\ 
\hline
\textbf{BFX}      & 446         & 0.542 & -0.032                       & 0.136 & -0.006                                   \\ 
\hline
\textbf{BSESN}    & 326         & 0.535 & -0.005                       & 0.104 & -0.007                                   \\ 
\hline
\textbf{BVLG}     & 461         & 0.472 & -0.023                       & 0.107 & -0.006                                   \\ 
\hline
\textbf{BVSP}     & 335         & 0.504 & -0.023                       & 0.109 & -0.004                                   \\ 
\hline
\textbf{DJI}      & 445         & 0.479 & -0.030                       & 0.105 & -0.005                                   \\ 
\hline
\textbf{FCHI}     & 578         & 0.499 & -0.032                       & 0.123 & -0.012                                   \\ 
\hline
\textbf{FTMIB}    & 256         & 0.485 & -0.026                       & 0.112 & -0.005                                   \\ 
\hline
\textbf{FTSE}     & 486         & 0.485 & -0.022                       & 0.105 & -0.009                                   \\ 
\hline
\textbf{GDAXI}    & 502         & 0.519 & -0.033                       & 0.131 & -0.011                                   \\ 
\hline
\textbf{GSPTSE}   & 337         & 0.512 & -0.022                       & 0.092 & -0.002                                   \\ 
\hline
\textbf{HSI}      & 669         & 0.503 & -0.024                       & 0.083 & -0.008                                   \\ 
\hline
\textbf{IBEX}     & 1070        & 0.520 & -0.030                       & 0.137 & -0.010                                   \\ 
\hline
\textbf{IXIC}     & 927         & 0.529 & -0.016                       & 0.103 & -0.009                                   \\ 
\hline
\textbf{KS11}     & 985         & 0.509 & -0.014                       & 0.088 & -0.007                                   \\ 
\hline
\textbf{KSE}      & 103         & 0.582 & -0.024                       & 0.112 & -0.000                                   \\ 
\hline
\textbf{MXX}      & 1096        & 0.540 & -0.039                       & 0.075 & -0.017                                   \\ 
\hline
\textbf{N225}     & 344         & 0.517 & -0.017                       & 0.096 & -0.003                                   \\ 
\hline
\textbf{NSEI}     & 477         & 0.531 & -0.014                       & 0.104 & -0.005                                   \\ 
\hline
\textbf{OMXC20}   & 439         & 0.517 & -0.020                       & 0.104 & -0.007                                   \\ 
\hline
\textbf{OMXHPI}   & 371         & 0.514 & -0.017                       & 0.117 & -0.008                                   \\ 
\hline
\textbf{OMXSPI}   & 444         & 0.506 & -0.015                       & 0.124 & -0.007                                   \\ 
\hline
\textbf{OSEAX}    & 332         & 0.531 & -0.014                       & 0.093 & -0.008                                   \\ 
\hline
\textbf{RUT}      & 229         & 0.470 & -0.005                       & 0.110 & -0.002                                   \\ 
\hline
\textbf{SMSI}     & 643         & 0.523 & -0.032                       & 0.128 & -0.012                                   \\ 
\hline
\textbf{SPX}      & 477         & 0.496 & -0.029                       & 0.115 & -0.002                                   \\ 
\hline
\textbf{SSEC}     & 511         & 0.570 & -0.020                       & 0.114 & -0.007                                   \\ 
\hline
\textbf{SSMI}     & 528         & 0.510 & -0.025                       & 0.135 & -0.004                                   \\ 
\hline
\textbf{STI}      & 598         & 0.557 & -0.018                       & 0.072 & -0.009                                   \\ 
\hline
\textbf{STOXX50E} & 504         & 0.504 & -0.035                       & 0.118 & -0.019                                   \\
\hline
\end{tabular}
}
\caption{Estimated maximum aggregation time, scaling, and multiscaling exponents for price and realized variance (at 10 minutes frequency) time series. All estimated values are statistically significant at $5\%$ confidence level.}\label{res_scaling}
\end{table}

After we have computed the Hurst exponent of the realized variance time series and the multiscaling proxy of the prices time series and confirmed that prices are indeed multiscaling, we proceed to compute the correlation coefficients in order to measure their interplay. We find a negative correlation between $\widehat{H}^{(v)}$ and $\widehat{B}^{(P)}$ using both the Pearson and Spearman correlations. We found a Pearson correlation coefficient of $-0.43$ and a Spearman correlation coefficient of $-0.51$, both statistically significant at $5\%$ level.

\subsection{Robust analysis}

\begin{figure}[H]
		\begin{center}	
			
			\includegraphics[width=1\textwidth,height=0.4\textheight]{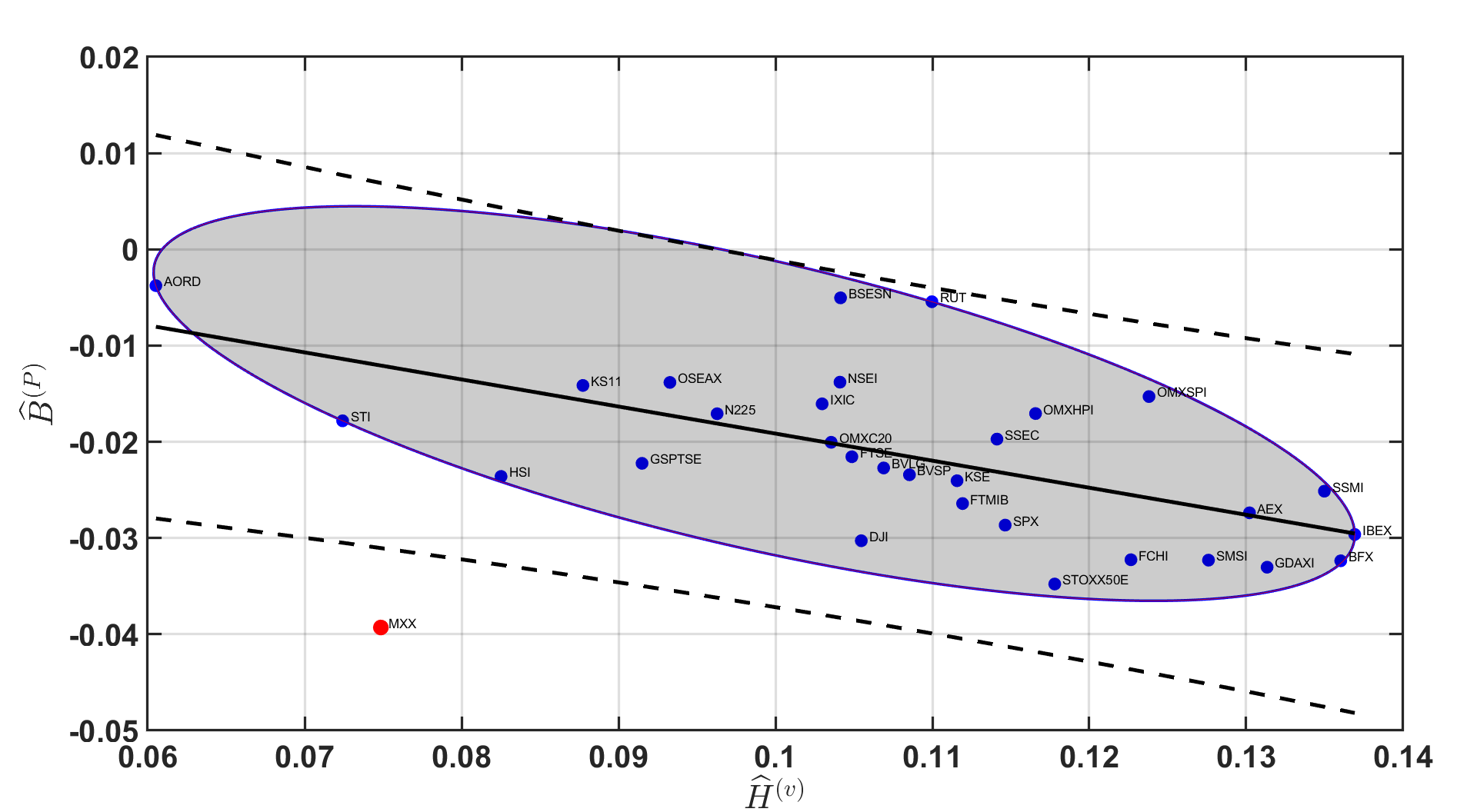} 
		\end{center}
		\caption{Estimated multiscaling proxy of the prices $\widehat{B}^{(P)}$ as function of volatility roughness $\widehat{H}^{(v)}$ (realized variance at 10 min frequency). Blue dots label different indices, the black continuous line is the regression line while the black dashed lines are the 95\% confidence intervals.  Pearson correlation coefficient $\rho$ is $-0.43$ and Spearman correlation coefficient $\rho_S$ is $-0.51$. The outlier-robust versions, $\tilde{\rho}$ and $\tilde{\rho}_S$ are equal to $-0.61$ and $-0.65$ respectively. All correlations are statistically significant at $5\%$ level.}
		\label{fig1}
	\end{figure}

Although easy to implement, Equation \ref{corr} is known to be strongly affected by outliers. In fact, even a very small portion of outliers can severely bias its estimated correlation coefficient. To tackle this issue, several methodologies have been proposed in the robust statistics literature. Some methodologies act at reducing the impact of the outliers by downweighting their contribution in the computation of the correlation coefficient while others methods compute the correlation over the outliers-filtered dataset. In a set of papers \citep{wilcox2004inferences,pernet2013robust,wilcox2018improved}, it has been shown that the second approach gives better results in reducing the bias. For this reason, we compute the correlation coefficient on the outlier filtered dataset. In particular, we are interested in removing the multivariate outliers, which are the relevant ones for the computation of the correlation coefficient. To this extent, we employ the bivariate outliers detection method of \citep{pernet2013robust}.\footnote{The entire procedure used to detect the outliers is reported in \ref{outliers_sec}.} From this analysis, we found out that IPC Mexico (MXX) is an outlier and for this reason, we label it as an outlier in our correlation analysis. We define the robust correlations as $\tilde{\rho}$ (or $\tilde{\rho_S}$), i.e. the correlation coefficient computed on the dataset without considering IPC Mexico (MXX). The results are depicted in Figure \ref{fig1}, where we also report the correlation coefficients that have not been corrected for the outlier.

As we can observe from Figure \ref{fig1}, the interplay between volatility roughness and prices' multiscaling is strongly negative. In fact, the rougher the volatility process is, the less multiscaling is the prices time series. Indeed, it is much stronger than what the found in the simulation experiment (for similar values of $H$) and the dependency is in the opposite direction. This result highlights the fact that the rBergomi model is not capable to reproduce this strong empirical dependency structure. This result calls for the development of models which are able to accommodate both multiscaling prices and rough volatility and their negative correlation. 

\begin{figure}[H]
		\begin{center}	
			
			\includegraphics[width=1\textwidth,height=0.45\textheight]{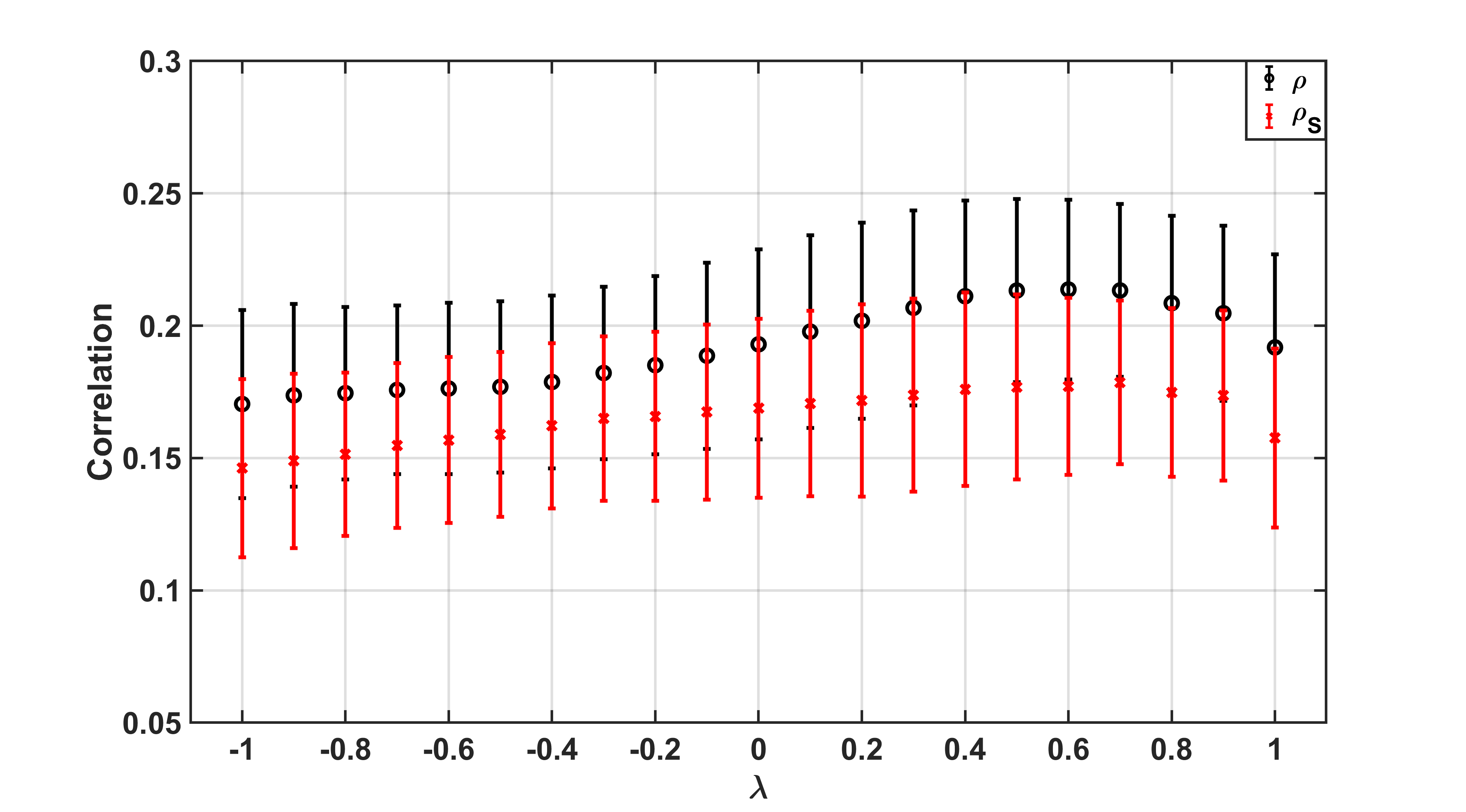} 
		\end{center}
		\caption{Correlation coefficients between $\widehat{H}^{(v)}$ and $\widehat{B}^{(P)}$ as function of $\lambda$ for the rBergomi model with $H$ taken from Table \ref{res_scaling}, $\xi=0.01$, $\eta=1.9$. Black line is the Pearson correlation while the red line corresponds to the Spearman correlation. Statistics computed over $100$ simulations, each composed by $31$ simulated paths each of $5000$ time steps. The error bars represent the standard errors. The plot has been smoothed via interpolation for better representation.}
		\label{fig1b}
	\end{figure}

Finally, to check if the result is dependent on the heterogeneity and distribution of the Hurst exponents $\widehat{H}^{(v)}$ of the real data and to the maximum time aggregation $\tau_{max}$ used for the estimation of the scaling exponents, we simulated $100$ set of time series (prices and volatility) using the rBergomi model, each one composed by $31$ time series with the specific set of $\widehat{H}^{(v)}$ estimated from the real data (see Table \ref{res_scaling}) and with a varying level of the correlation parameter $\lambda$ (the remaining parameters are left unchanged). The estimation is then carried over by using the same procedure described in Section \ref{res_procedure}, by using the same $\tau_{max}$ of Table \ref{res_scaling}. Results are reported in Figure \ref{fig1b}, where we report both correlation measures as a function of $\lambda$.\footnote{Additional results related to the correlation between other scaling measures are reported in \ref{add_res_real}.}

As it is possible to observe, the correlation is positive for each value of $\lambda$ with a peak near the positive boundary. This highlights the fact that it is not the heterogeneity of the true data that produced the result, which is indeed robust. 

\section{Summary and final remarks}\label{sec_c}
To check for any interplay between prices' multiscaling and volatility roughness, we have produced extensive simulation experiments by using one of the benchmark models in the financial mathematics literature on rough volatility, namely the rough Bergomi model. By using the model parameters in \citep{rBergomi,rBergomi_calibration} and by changing the Wiener processes correlation and the Hurst exponent, we have investigated if the simulated volatility and price processes showed any relationship in their scaling exponents. We have found that the correlation between prices' multiscaling and rough volatility is mainly positive, peaking for small values of $H$, while the correlation parameter $\lambda$ does not play a major role in this relationship.
We have then computed the same dependency measures by using real data. We have found that there exists a statistically significant (negative) dependence between volatility roughness and prices' multiscaling by analysing different indices. In particular, we have found that the rougher the volatility is, the less multiscaling the price series are. This result shows that even if the rBergomi is able to produce multiscaling prices for low values of $H$, the empirical dependence is reversed. To check if the heterogeneity of the empirical scaling exponents was producing an artifact dependency structure, we have produced a new simulation experiment in which the scaling exponents of the volatility process was taken from the ones we have estimated from the real data. Even in this case, we found that the model is not able to reproduce the interplay found in the real data. This result shows that current models are not able to reproduce this higher-order dependence between the scaling features of the volatility and prices processes. Indeed, calibrating $H$ in the volatility process, would produce the opposite effect on the prices multiscaling, e.g. low $H$ would imply higher multiscaling prices while it should generate prices with low level of multiscaling.
A possible solution to this is to employ multiscaling models for the price's fluctuations and a fractional type of process for the volatility dynamics. In particular, it would be advisable to link the multiscaling measures of the prices' process with the volatility roughness. One possibility would be to implement a time-changed Brownian motion for the log-prices fluctuations, where the time change measure is indeed multifractal with the intermittency parameter linked to the Hurst exponent of the underlying volatility process. This will generate more reliable price time series that in combination with turbo-charged Monte Carlo procedures \citep{mccrickerd2018turbocharging} can be used to make forecasts and price Options.
Future analysis might include the investigation of the dynamic dependency between the scaling measures in order to check for trends and cycles. Finally, since rough volatility is not directly observed but proxied by various measures (realized variance for example), it would be beneficial to understand the impact of such volatility proxies on the dependency structure between prices multiscaling and volatility roughness.

\bibliographystyle{plainnat}

\bibliography{mulitscaling_and_rough_volatility}







\newpage
\appendix
\renewcommand{\thesubsection}{\thesection.\Roman{subsection}}

\section{Additional results: Synthetic data}\label{add_res_sin}

\setcounter{table}{0}
\setcounter{figure}{0}
In this Section, we report additional results related to the analysis of scaling exponents of both volatility and prices processes and their interplay related to the synthetic data.

		\begin{figure}[H]
		\begin{center}	
			\includegraphics[width=0.9\textwidth,height=0.34\textheight]{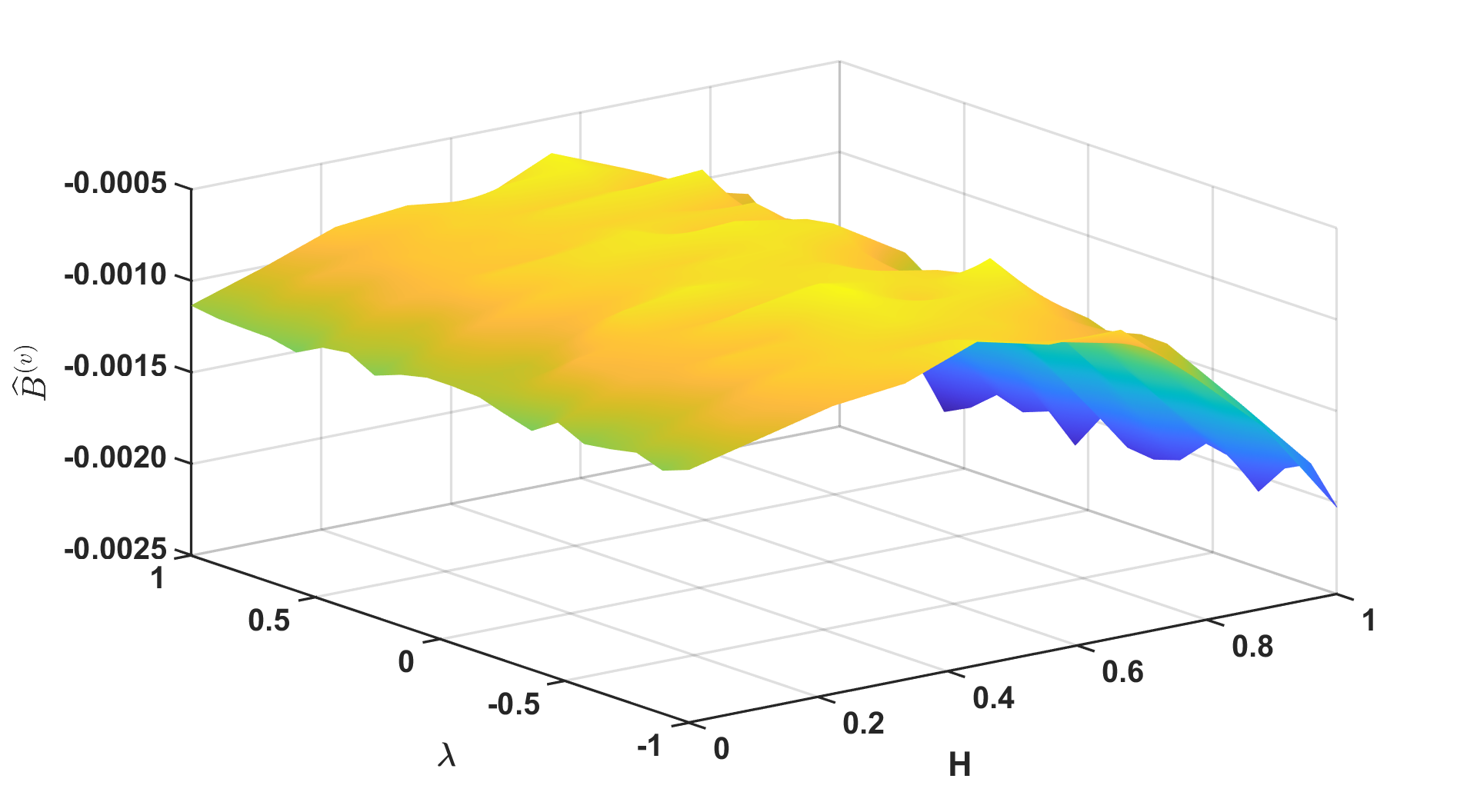} 
		\end{center}
		\caption{Multiscaling proxy $\widehat{B}^{(v)}$ with respect to $H$ and $\lambda$ in the rBergomi model. The result is averaged over the $100$ dataset and the plot is smoothed via interpolation for better representation.}
		\label{fig_apx1}
	\end{figure} 

As it is possible to see from Figure \ref{fig_apx1}, even if the level of multiscaling increases with respect to $H$, it remains negligible also for $H\sim1$.

	\begin{figure}[H]
		\begin{center}	
			\includegraphics[width=0.9\textwidth,height=0.34\textheight]{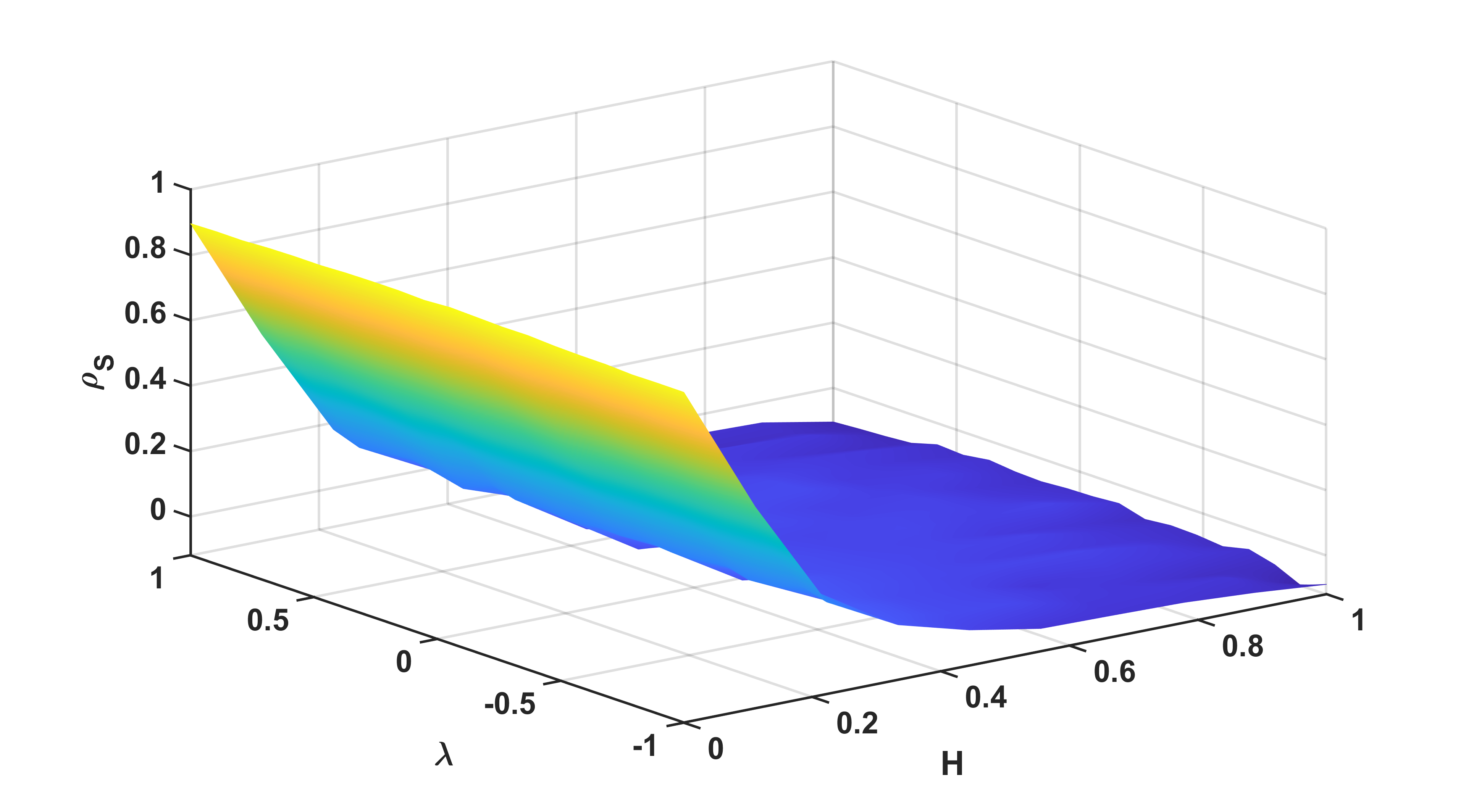} 
		\end{center}
		\caption{Spearman correlation $\rho_S$ between the multiscaling proxy $\widehat{B}^{(P)}$ and $\widehat{H}^{(v)}$. Description as for caption of Figure \ref{fig6a}.}
		\label{fig_apx2}
	\end{figure} 
As for the Pearson correlation of Figure \ref{fig6a}, we find that the Spearman correlation between the multiscaling proxy $\widehat{B}^{(P)}$ and $\widehat{H}^{(v)}$ is higher for small values of $H$ and becomes negligible for $H>0.3$. \par

For completeness, we also report the correlations between the Hurst exponent of the prices and volatility processes as well as the dependence between their multiscaling features.

		\begin{figure}[H]
		\begin{center}	
			\includegraphics[width=0.9\textwidth,height=0.34\textheight]{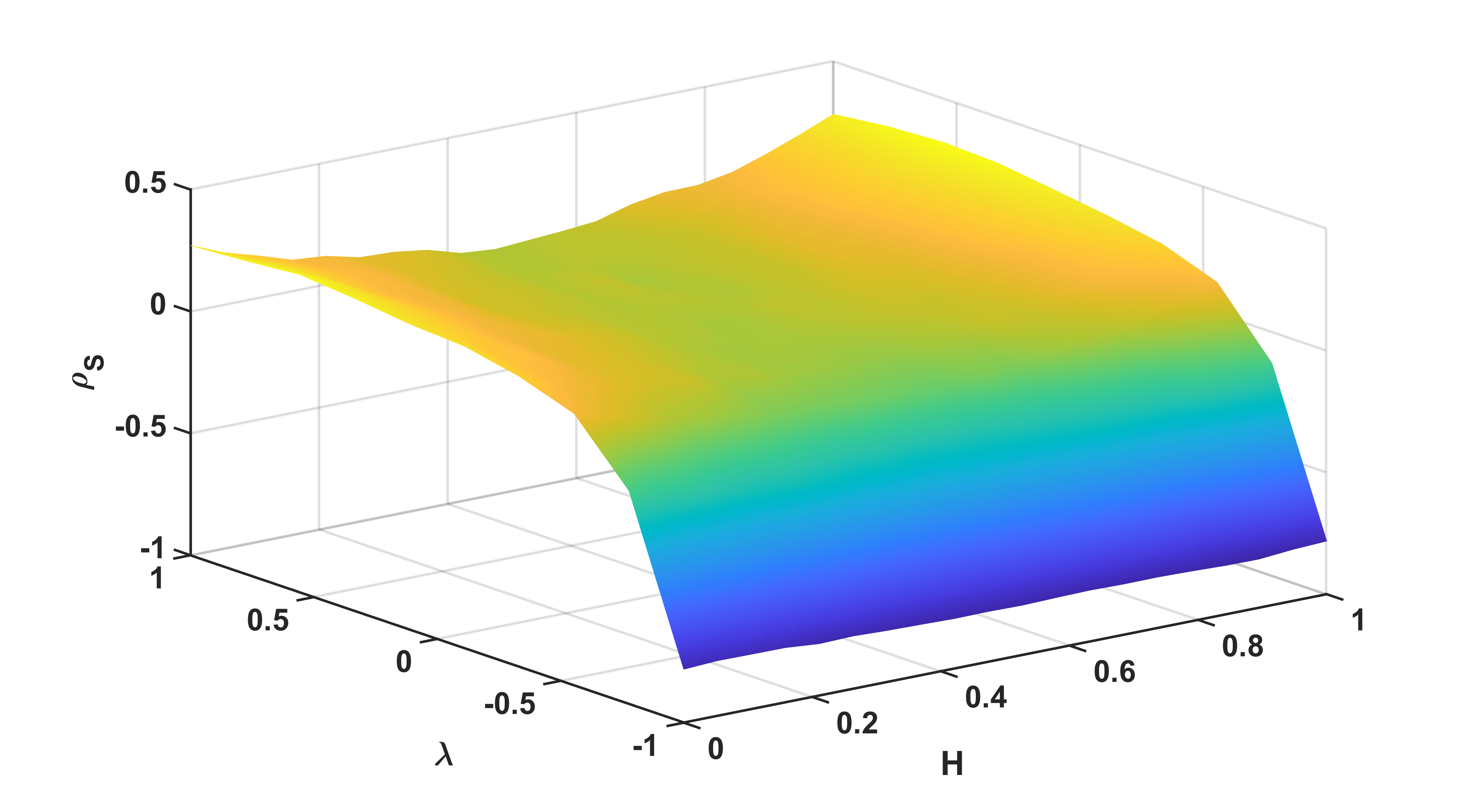} 
		\end{center}
		\caption{Spearman correlation between Multiscaling proxy $\widehat{H}^{(P)}$ and $\widehat{H}^{(v)}$. Description as for caption of Figure \ref{fig6a}.}
		\label{fig_apx4}
	\end{figure} 

As we can observe from Figure \ref{fig_apx4}, the correlation between the Hurst exponents of the two processes is almost entirely generated by the correlation parameter $\lambda$. This is not an unexpected result since the $\lambda$ drives the correlation of the diffusive components of the two processes.

	\begin{figure}[H]
		\begin{center}	
			\includegraphics[width=0.9\textwidth,height=0.34\textheight]{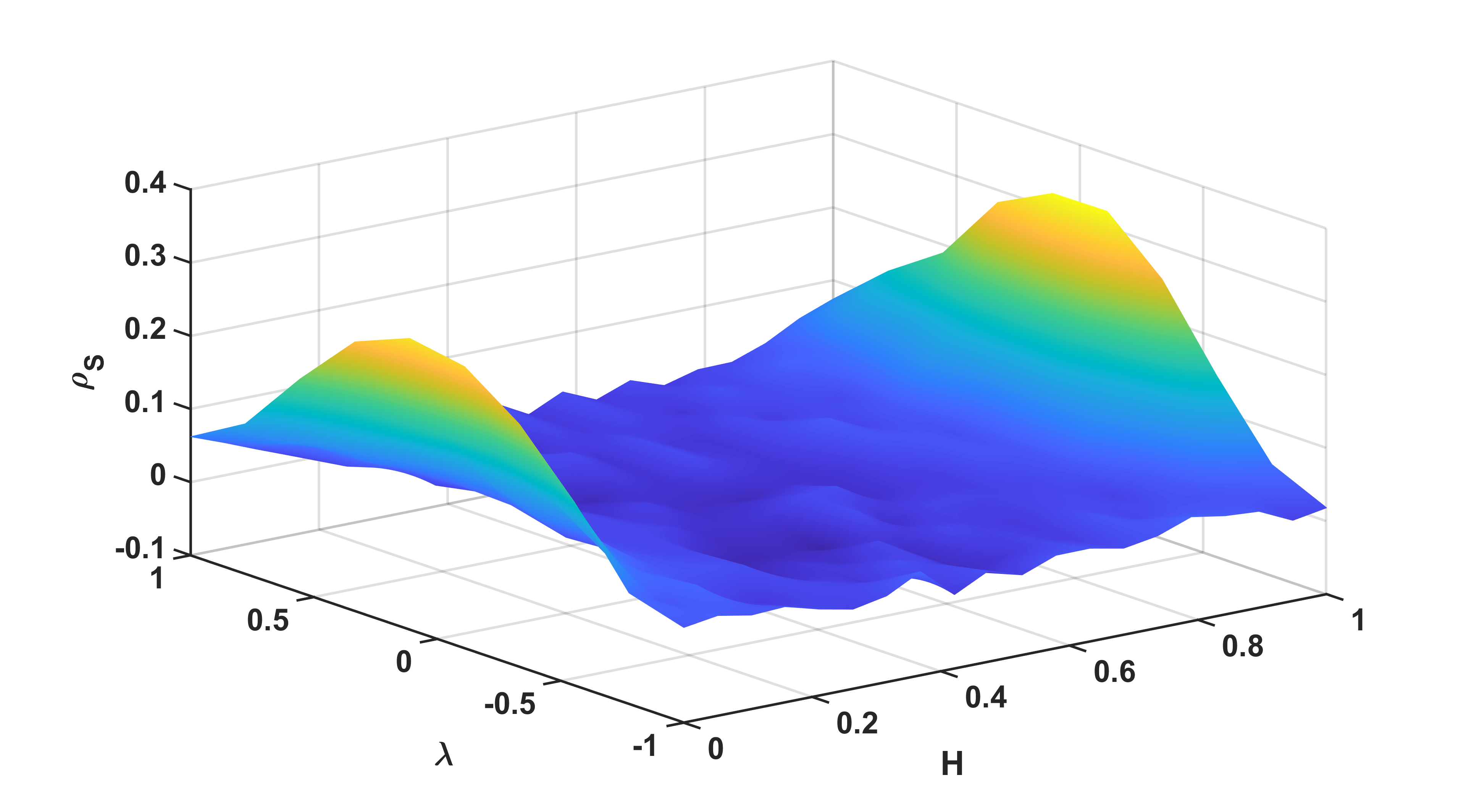} 
		\end{center}
		\caption{Spearman correlation between Multiscaling proxy $\widehat{B}^{(P)}$ and $\widehat{B}^{(v)}$. Description as for caption of Figure \ref{fig6a}.}
		\label{fig_apx3}
	\end{figure} 
	
With respect to the interplay between the multiscaling features of the two processes, there is a correlation at the boundaries of the parameter $H$. This is due to the following motivation. As it is possible to see from Figure \ref{fig5c}, the slope of the prices' multiscaling proxy $\widehat{B}^{(P)}$ is strongly positive for $H\sim0$, then it becomes flat for intermediate values of $H$. For high values of $H$, even if not statistically significant, it becomes slightly negative. The same type of behavior, even if with a different strength, is reported in Figure \ref{fig_apx1} for the multiscaling feature of the volatility process. Indeed, the slope of the of $\widehat{B}^{(v)}$ for small values of $H$, even if not statistically significant, is positive, while is negative for $H\sim1$. For these reasons, the correlation between the multiscaling features of the two processes is positive at the boundaries of $H$. However, these correlations are not statistically significant at $5\%$ level.     

\newpage

\section{Data}\label{sec_data}
\setcounter{table}{0}
\setcounter{figure}{0}
The data used in this paper are taken from the Oxford volatility library \citep{library}. Codes and descriptions are reported in the table \ref{Tab_library} while the stock indices available and the time periods for which the data are available are reported in Table \ref{Tab_indices}. The data is checked for missing values and in the cases in which a data-point is not available, a linear interpolation method is used to input the datum.

\begin{table}[h]
\centering
\begin{tabular}{|l|l|}
\hline
Code         & Description                      \\
\hline
close\_price & Closing (Last) Price             \\
\hline
open\_to\_close & Open to Close log-Return      \\
\hline
rv10         & Realized Variance (10-min)       \\
\hline
rv5          & Realized Variance (5-min)        \\
\hline
rsv          & Realized Semi-variance (5-min) \\
\hline
bv           & Bipower Variation (5-min)   \\
\hline    
\end{tabular}
\caption{Variables of the Oxford Volatility Library used in the paper.}\label{Tab_library}
\end{table}

\textbf{close\_price:} Daily closing price. The closing price is the last observed price of the day.\\

\textbf{open\_to\_close:} Daily open to close returns are the log-returns computed between the opening price and the closing price for each day.\\

\textbf{rv5 and rv10:} Realized variance at 5 minutes and 10 minutes sampling frequency. These measures are computed as the sum of squared returns over a specific time window and a specific time frequency. For example, the RV can be the sum of squared intra-day returns at 10 minutes frequency.\\

\textbf{rsv5:} Realized semi-variance at 5 minutes sampling frequency. The realized semi-variance is calculated by first computing the realized variance for negative and positive returns separately and then summing them up.\\

\textbf{bv:} Realized Bipower Variation at 5 minutes sampling frequency. Bipower variation is computed as the (scaled) sum of products of adjacent absolute returns.

\begin{table}[H]
\resizebox{\textwidth}{!}{%
\begin{tabular}{|l|l|l|l|}
\hline
\textbf{Index} & \textbf{\textbf{Market name}} & \textbf{\textbf{First date}} & \textbf{\textbf{Last date}} \\ \hline
\textbf{AEX}      & AEX index                         & 03/01/2000 & 11/11/2021 \\ \hline
\textbf{AORD}     & All Ordinaries                    & 04/01/2000 & 11/11/2021 \\ \hline
\textbf{BFX}      & Bell 20 Index                     & 03/01/2000 & 11/11/2021 \\ \hline
\textbf{BSESN}    & S\&P BSE Sensex                   & 03/01/2000 & 11/11/2021 \\ \hline
\textbf{BVLG}     & PSI All-Share Index               & 15/10/2012 & 11/11/2021 \\ \hline
\textbf{BVSP}     & BVSP BOVESPA Index                & 03/01/2000 & 11/11/2021 \\ \hline
\textbf{DJI}      & Dow Jones Industrial   Average    & 03/01/2000 & 11/11/2021 \\ \hline
\textbf{FCHI}     & CAC 40                            & 03/01/2000 & 11/11/2021 \\ \hline
\textbf{FTMIB}    & FTSE MIB                          & 01/06/2009 & 11/11/2021 \\ \hline
\textbf{FTSE}     & FTSE 100                          & 04/01/2000 & 11/11/2021 \\ \hline
\textbf{GDAXI}    & DAX                               & 03/01/2000 & 11/11/2021 \\ \hline
\textbf{GSPTSE}   & S\&P/TSX Composite index          & 02/05/2002 & 11/11/2021 \\ \hline
\textbf{HSI}      & HANG SENG Index                   & 03/01/2000 & 11/11/2021 \\ \hline
\textbf{IBEX}     & IBEX 35 Index                     & 03/01/2000 & 11/11/2021 \\ \hline
\textbf{IXIC}     & Nasdaq 100                        & 03/01/2000 & 11/11/2021 \\ \hline
\textbf{KS11}     & Korea Composite Stock Price Index & 04/01/2000 & 11/11/2021 \\ \hline
\textbf{KSE}      & Karachi SE 100 Index              & 03/01/2000 & 11/11/2021 \\ \hline
\textbf{MXX}      & IPC Mexico                        & 03/01/2000 & 11/11/2021 \\ \hline
\textbf{N225}     & Nikkei 225                        & 02/02/2000 & 11/11/2021 \\ \hline
\textbf{NSEI}     & NIFTY 50                          & 03/01/2000 & 11/11/2021 \\ \hline
\textbf{OMXC20}   & OMX Copenhagen 20 Index           & 03/10/2005 & 11/11/2021 \\ \hline
\textbf{OMXHPI}   & OMX Helsinki All Share   Index    & 03/10/2005 & 11/11/2021 \\ \hline
\textbf{OMXSPI}   & OMX Stockholm All Share   Index   & 03/10/2005 & 11/11/2021 \\ \hline
\textbf{OSEAX}    & Oslo Exchange All-share   Index   & 03/09/2001 & 11/11/2021 \\ \hline
\textbf{RUT}      & Russel 2000                       & 03/01/2000 & 11/11/2021 \\ \hline
\textbf{SMSI}     & Madrid General Index              & 04/07/2005 & 11/11/2021 \\ \hline
\textbf{SPX}      & S\&P 500 Index                    & 03/01/2000 & 11/11/2021 \\ \hline
\textbf{SSEC}     & Shanghai Composite Index          & 04/01/2000 & 11/11/2021 \\ \hline
\textbf{SSMI}     & Swiss Stock Market Index          & 04/01/2000 & 11/11/2021 \\ \hline
\textbf{STI}      & Straits Times Index               & 03/01/2000 & 11/11/2021 \\ \hline
\textbf{STOXX50E} & EURO STOXX 50                     & 03/01/2000 & 11/11/2021 \\ \hline
\end{tabular}%
}
\caption{Information of the Oxford volatility library dataset.}\label{Tab_indices}
\end{table}

\newpage

\section{Additional results: Real data}\label{add_res_real}
\setcounter{table}{0}
\setcounter{figure}{0}
In this section, we report the additional results related to the analysis of the real data. When computing the correlation over the real data scaling features, we report both the standard and robust measures of correlations. 
\subsection{Volatility roughness and price multiscaling}
In this subsection, we report additional results related to different volatility measures with respect to the one presented in the main text.

	\begin{figure}[H]
		\begin{center}	
			
			\includegraphics[width=1\textwidth,height=0.35\textheight]{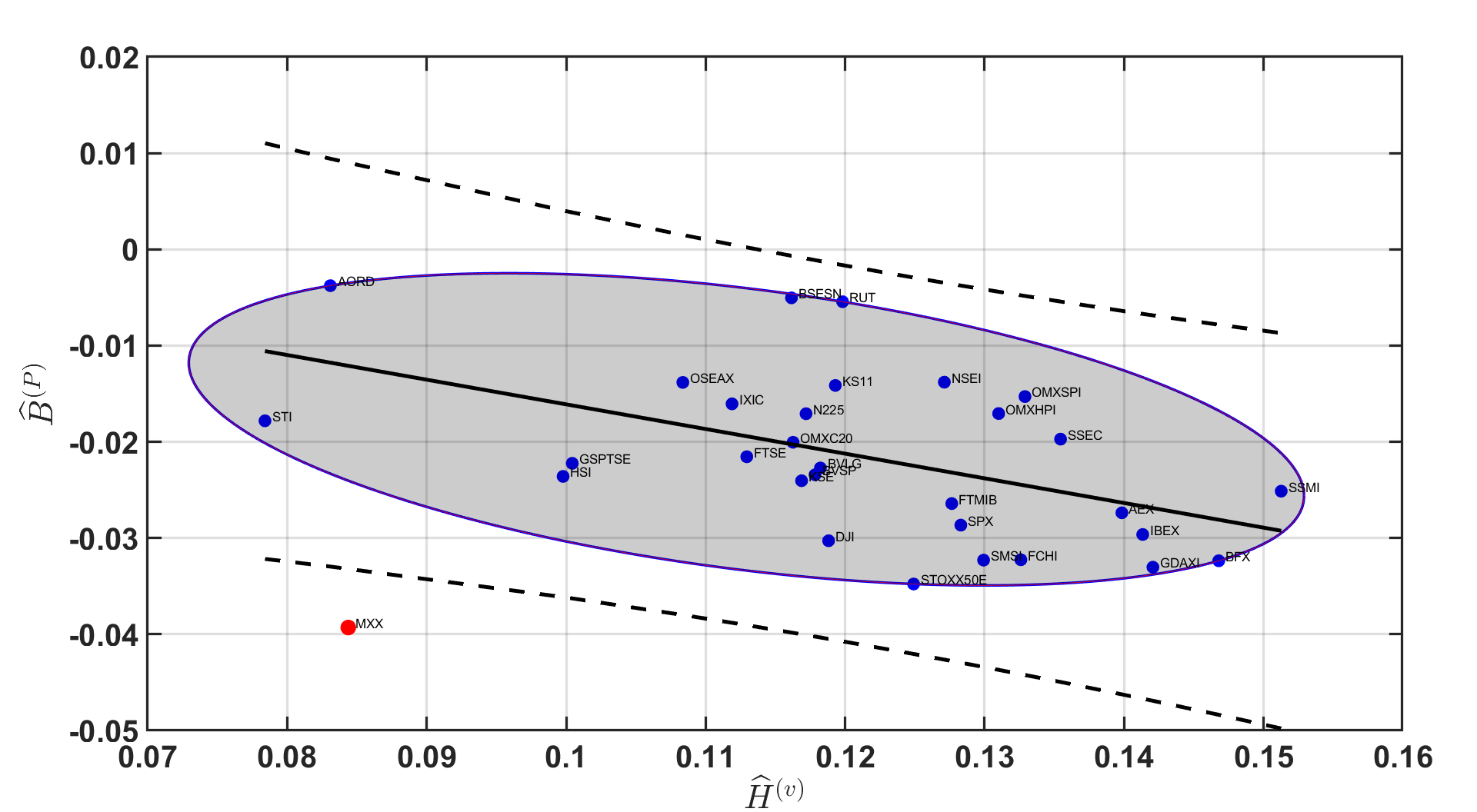} 
		\end{center}
		\caption{Estimated multiscaling proxy of the prices $\widehat{B}^{(P)}$ as function of volatility roughness $\widehat{H}^{(v)}$ (realized variance at 5 min frequency). Pearson correlation coefficient $\rho$ is $-0.30$ and Spearman correlation coefficient $\rho_S$ is $-0.39$. The outlier-robust versions, $\tilde{\rho}$ and $\tilde{\rho}_S$ are equal to $-0.51$ and $-0.51$ respectively. Description as in caption of Figure \ref{fig1}.}
		\label{fig_apx5}
	\end{figure}

		\begin{figure}[H]
		\begin{center}	
			
			\includegraphics[width=1\textwidth,height=0.35\textheight]{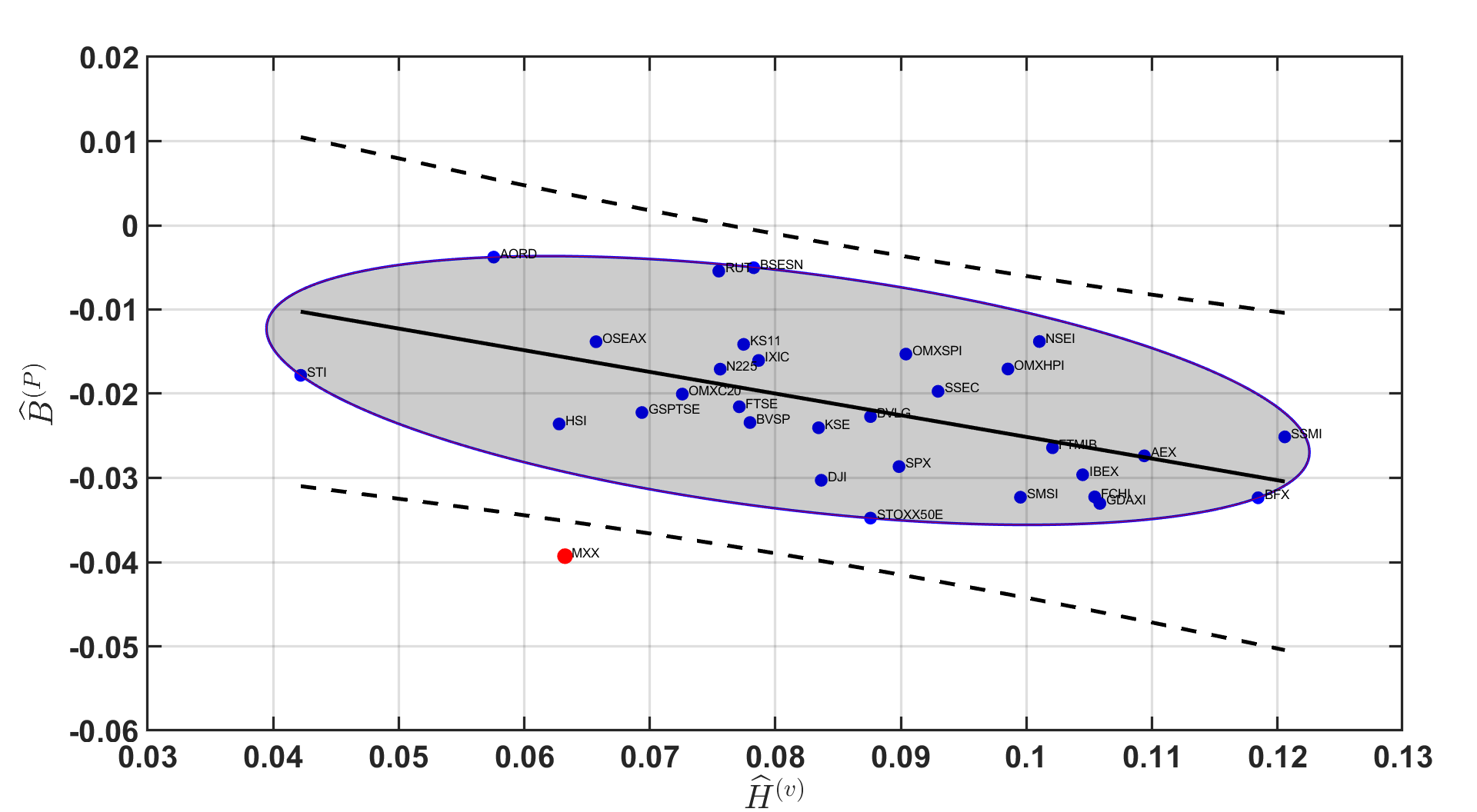} 
		\end{center}
		\caption{Estimated multiscaling proxy of the prices $\widehat{B}^{(P)}$ as function of volatility roughness $\widehat{H}^{(v)}$ (realized semi-variance at 5 min frequency). Pearson correlation coefficient $\rho$ is $-0.42$ and Spearman correlation coefficient $\rho_S$ is $-0.45$. The outlier-robust versions, $\tilde{\rho}$ and $\tilde{\rho}_S$ are equal to $-0.55$ and $-0.57$ respectively. Description as in caption of Figure \ref{fig1}.}
\label{fig_apx6}
\end{figure}

		\begin{figure}[H]
		\begin{center}	
			
			\includegraphics[width=1\textwidth,height=0.35\textheight]{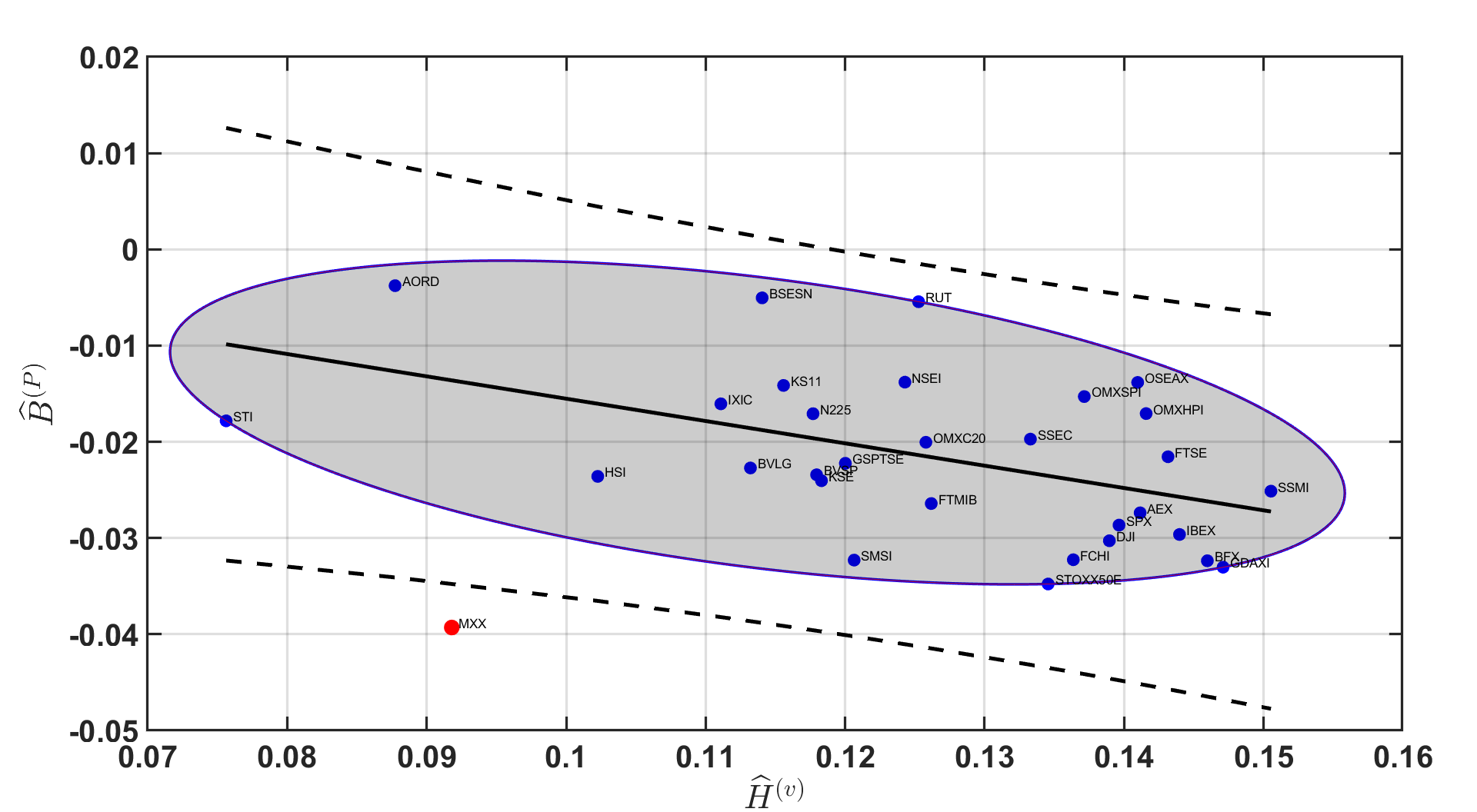} 
		\end{center}
		\caption{Estimated multiscaling proxy of the prices $\widehat{B}^{(P)}$ as function of volatility roughness $\widehat{H}^{(v)}$ (bipower variation at 5 min frequency). Pearson correlation coefficient $\rho$ is $-0.30$ and Spearman correlation coefficient $\rho_S$ is $-0.36$. The outlier-robust versions, $\tilde{\rho}$ and $\tilde{\rho}_S$ are equal to $-0.48$ and $-0.48$ respectively. Description as in caption of Figure \ref{fig1}.}
\label{fig_apx7}
\end{figure}

As we can see from the figures, the same pattern as for the realized variance at 10 minutes sampling frequency analyzed in the main text is retrieved for different rough volatility measures.

\subsection{Volatility roughness and price scaling}
In this subsection, we report additional results related to the study of the correlation between volatility roughness and price scaling for different volatility measures.
	\begin{figure}[H]
		\begin{center}	
			
			\includegraphics[width=1\textwidth,height=0.35\textheight]{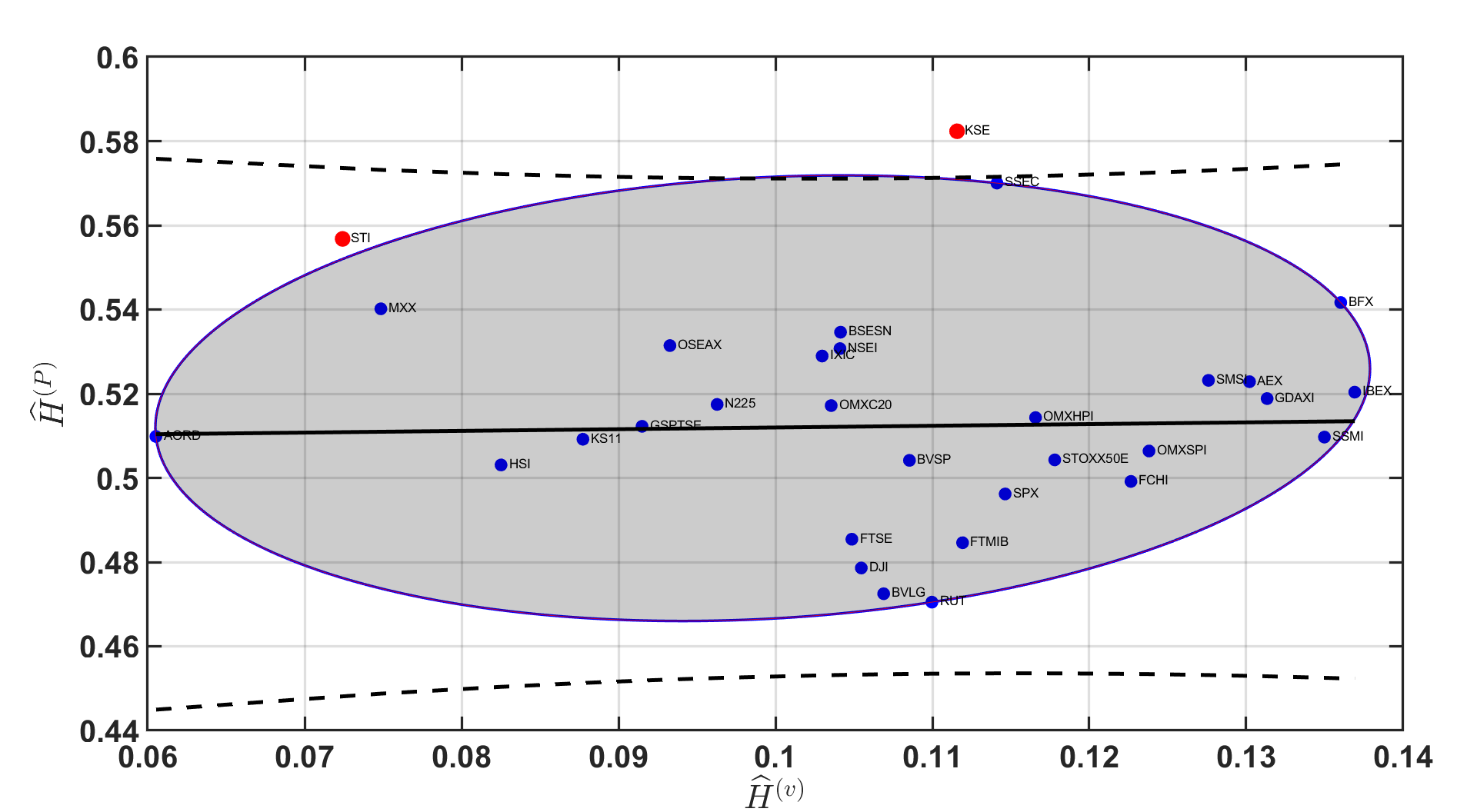} 
		\end{center}
		\caption{Estimated Hurst exponent of the prices $\widehat{H}^{(P)}$ as function of volatility roughness $\widehat{H}^{(v)}$ (realized variance at 10 min frequency). Pearson correlation coefficient $\rho$ is $-0.06$ and Spearman correlation coefficient $\rho_S$ is $-0.06$. The outlier-robust versions, $\tilde{\rho}$ and $\tilde{\rho}_S$ are equal to $0.03$ and $0.00$ respectively. Description as in caption of Figure \ref{fig1}.}
\label{fig_apx8}
\end{figure}

	\begin{figure}[H]
		\begin{center}	
			
			\includegraphics[width=1\textwidth,height=0.35\textheight]{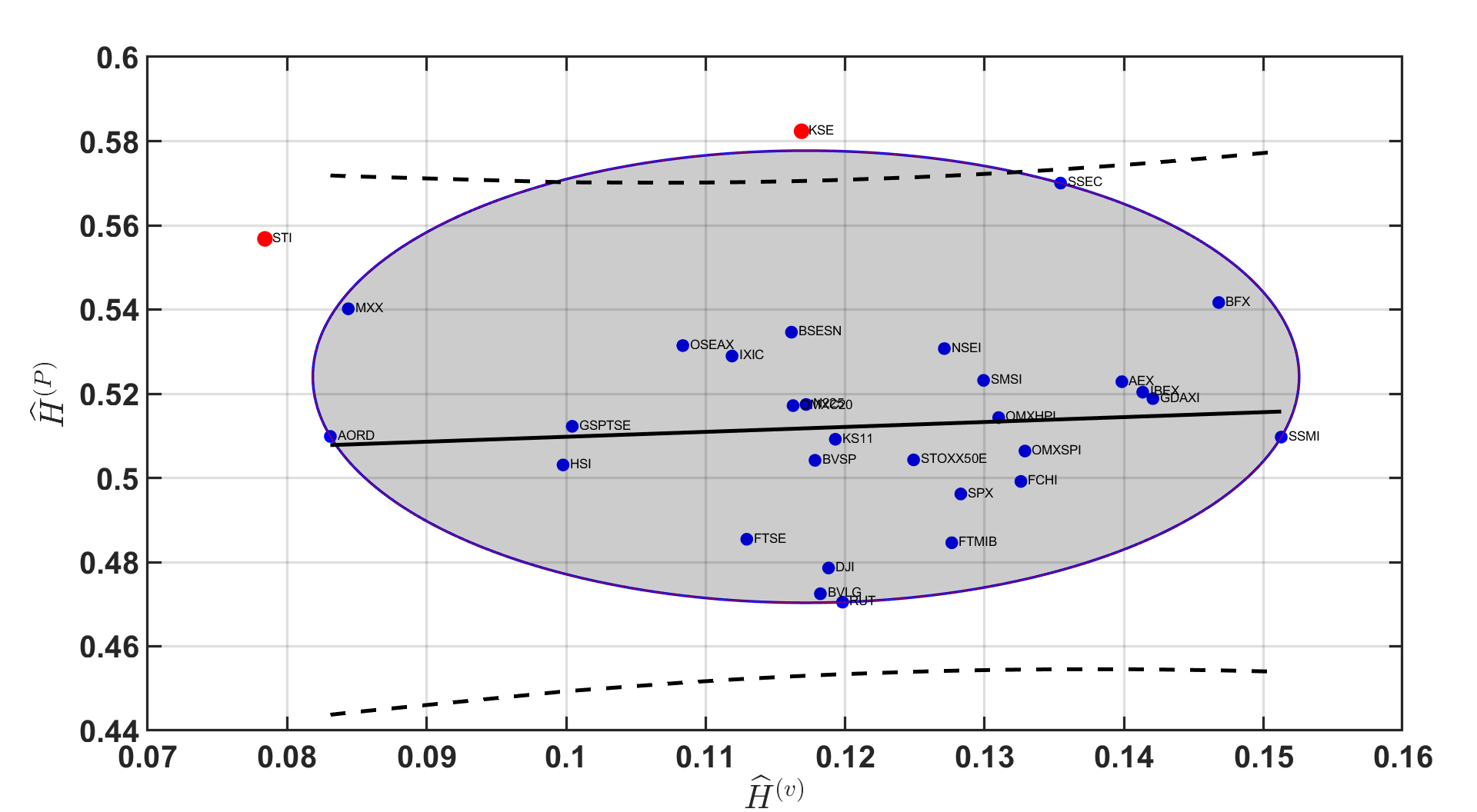} 
		\end{center}
		\caption{Estimated Hurst exponent of the prices $\widehat{H}^{(P)}$ as function of volatility roughness $\widehat{H}^{(v)}$ (realized variance at 5 min frequency). Pearson correlation coefficient $\rho$ is $-0.09$ and Spearman correlation coefficient $\rho_S$ is $-0.06$. The outlier-robust versions, $\tilde{\rho}$ and $\tilde{\rho}_S$ are equal to $0.09$ and $0.08$ respectively. Description as in caption of Figure \ref{fig1}.}
\label{fig_apx9}
\end{figure}

	\begin{figure}[H]
		\begin{center}	
			
			\includegraphics[width=1\textwidth,height=0.35\textheight]{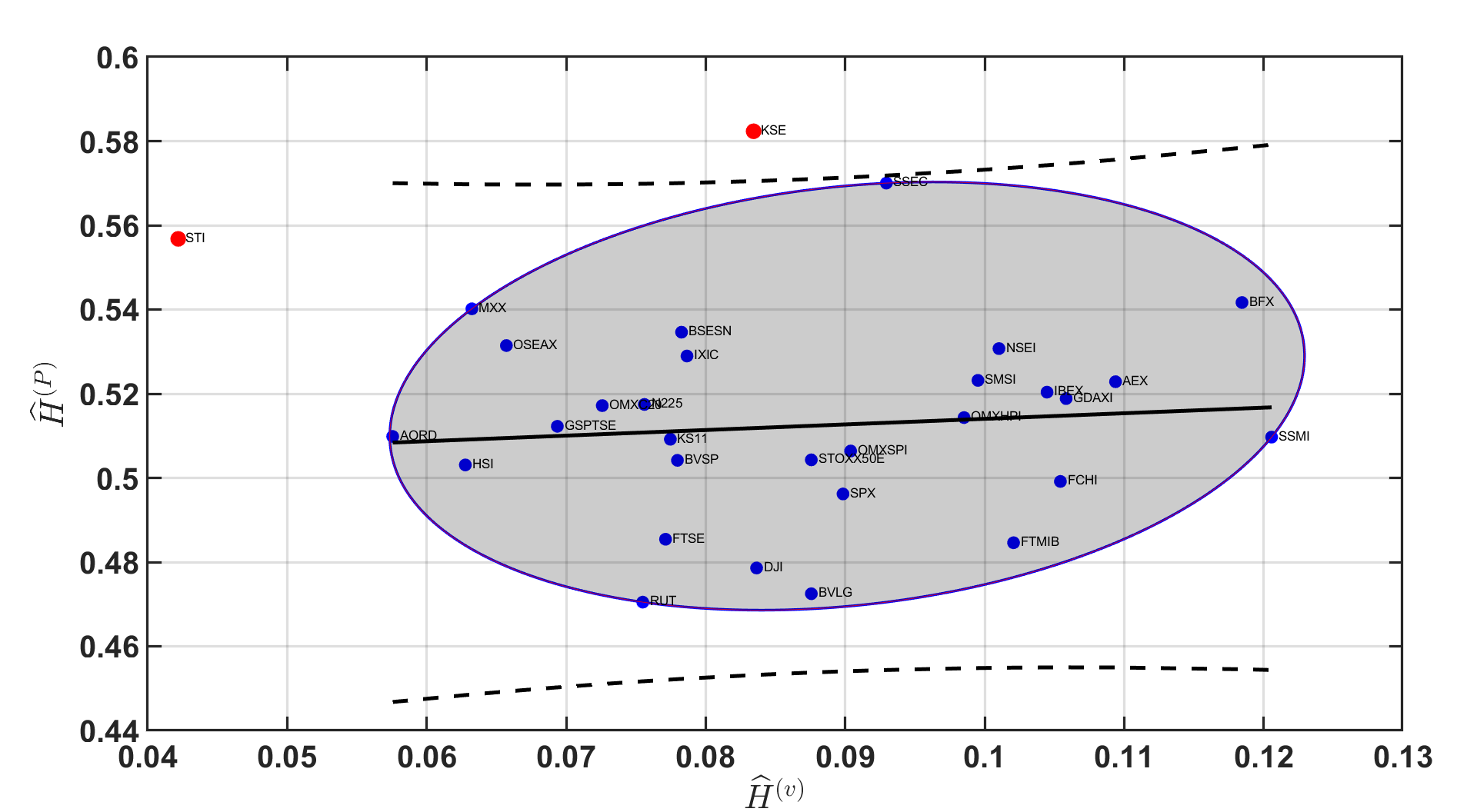} 
		\end{center}
		\caption{Estimated Hurst exponent of the prices $\widehat{H}^{(P)}$ as function of volatility roughness $\widehat{H}^{(v)}$ (realized semi-variance at 5 min frequency). Pearson correlation coefficient $\rho$ is $-0.07$ and Spearman correlation coefficient $\rho_S$ is $0.01$. The outlier-robust versions, $\tilde{\rho}$ and $\tilde{\rho}_S$ are equal to $0.10$ and $0.12$ respectively. Description as in caption of Figure \ref{fig1}.}
		\label{fig_apx10}
	\end{figure}

	\begin{figure}[H]
		\begin{center}	
			
			\includegraphics[width=1\textwidth,height=0.35\textheight]{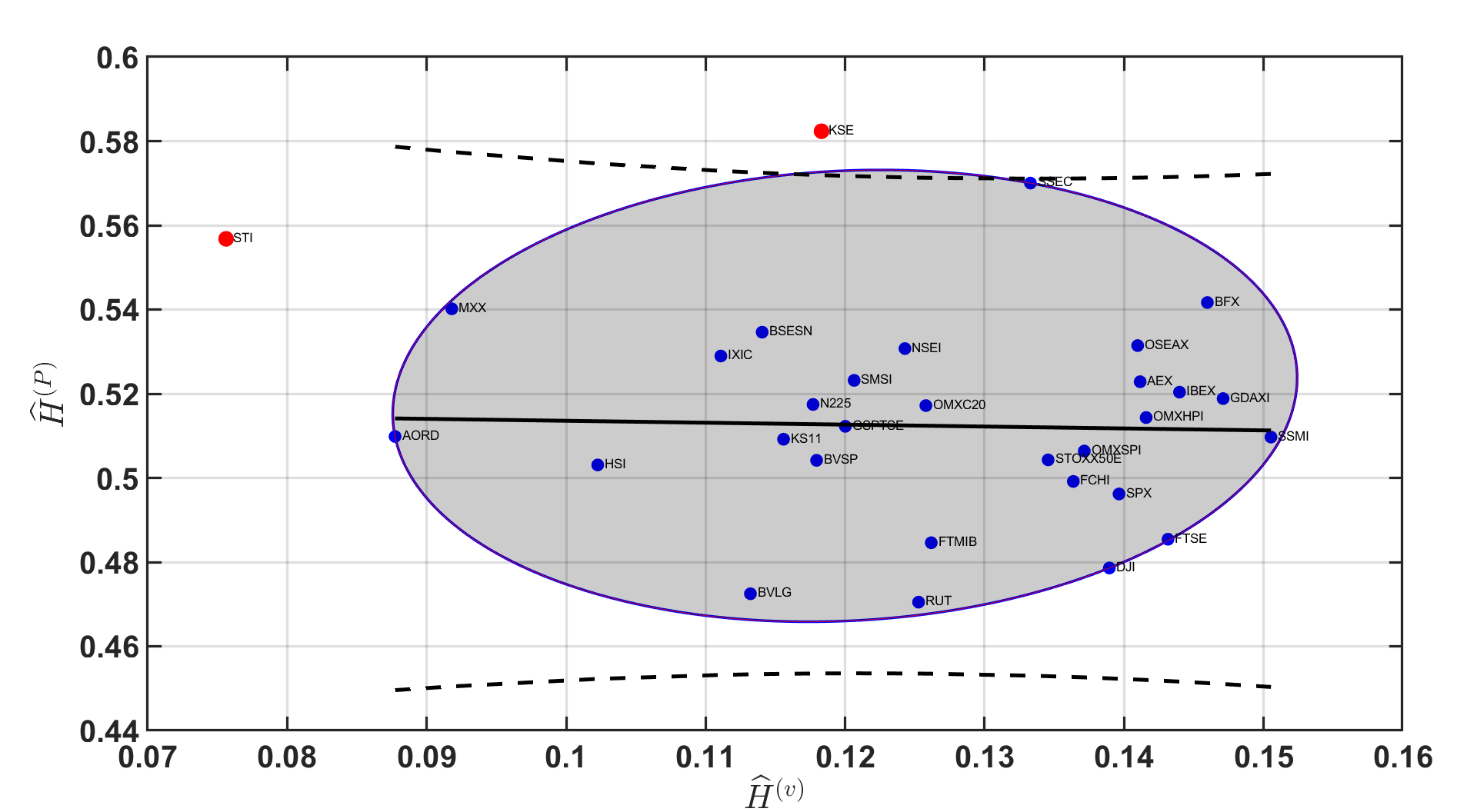} 
		\end{center}
		\caption{Estimated Hurst exponent of the prices $\widehat{H}^{(P)}$ as function of volatility roughness $\widehat{H}^{(v)}$ (bipower variation at 5 min frequency). Pearson correlation coefficient $\rho$ is $-0.21$ and Spearman correlation coefficient $\rho_S$ is $-0.11$. The outlier-robust versions, $\tilde{\rho}$ and $\tilde{\rho}_S$ are equal to $-0.03$ and $0.00$ respectively. Description as in caption of Figure \ref{fig1}.}
		\label{fig_apx11}
	\end{figure}

As it is possible to notice from the plots of this section, there is no statistical relationship between the Hurst exponents of the volatility and prices time series. Indeed, all correlation coefficients are not statistically significant. This result is confirmed also for the robust correlation coefficients.

\subsection{Volatility multiscaling and price multiscaling}

In this subsection, we report additional results related to the study of the correlation between volatility multiscaling and prices multiscaling for different volatility measures.

	\begin{figure}[H]
		\begin{center}	
			
			\includegraphics[width=1\textwidth,height=0.35\textheight]{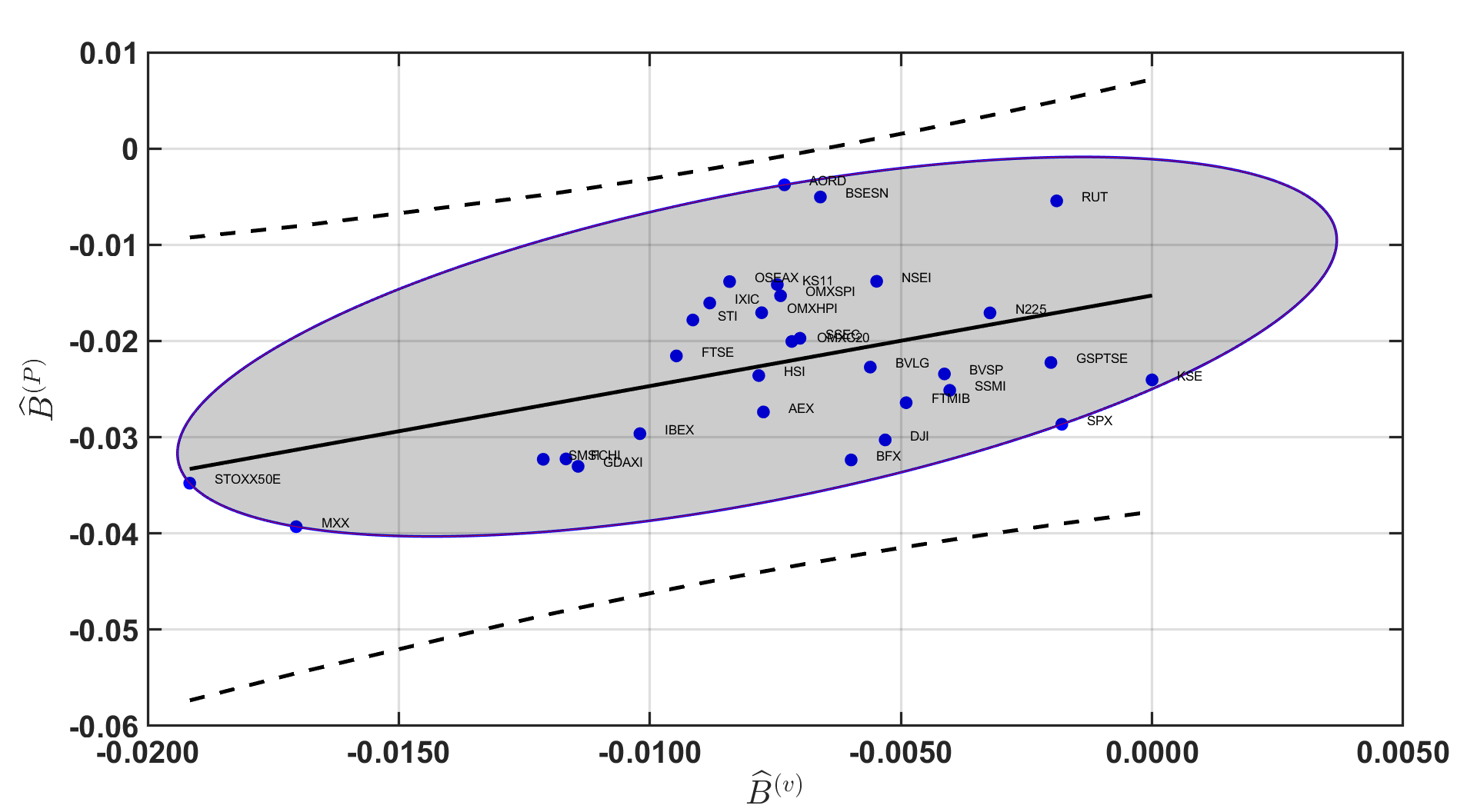} 
		\end{center}
		\caption{Estimated multiscaling proxy of the prices $\widehat{B}^{(P)}$ as function of volatility multiscaling $\widehat{B}^{(v)}$ (realized variance at 10 min frequency). Pearson correlation coefficient $\rho$ is $0.44$ and Spearman correlation coefficient $\rho_S$ is $0.29$. No outlier has been detected. Description as in caption of Figure \ref{fig1}.}
\label{fig_apx12}
\end{figure}

	\begin{figure}[H]
		\begin{center}	
			
			\includegraphics[width=1\textwidth,height=0.35\textheight]{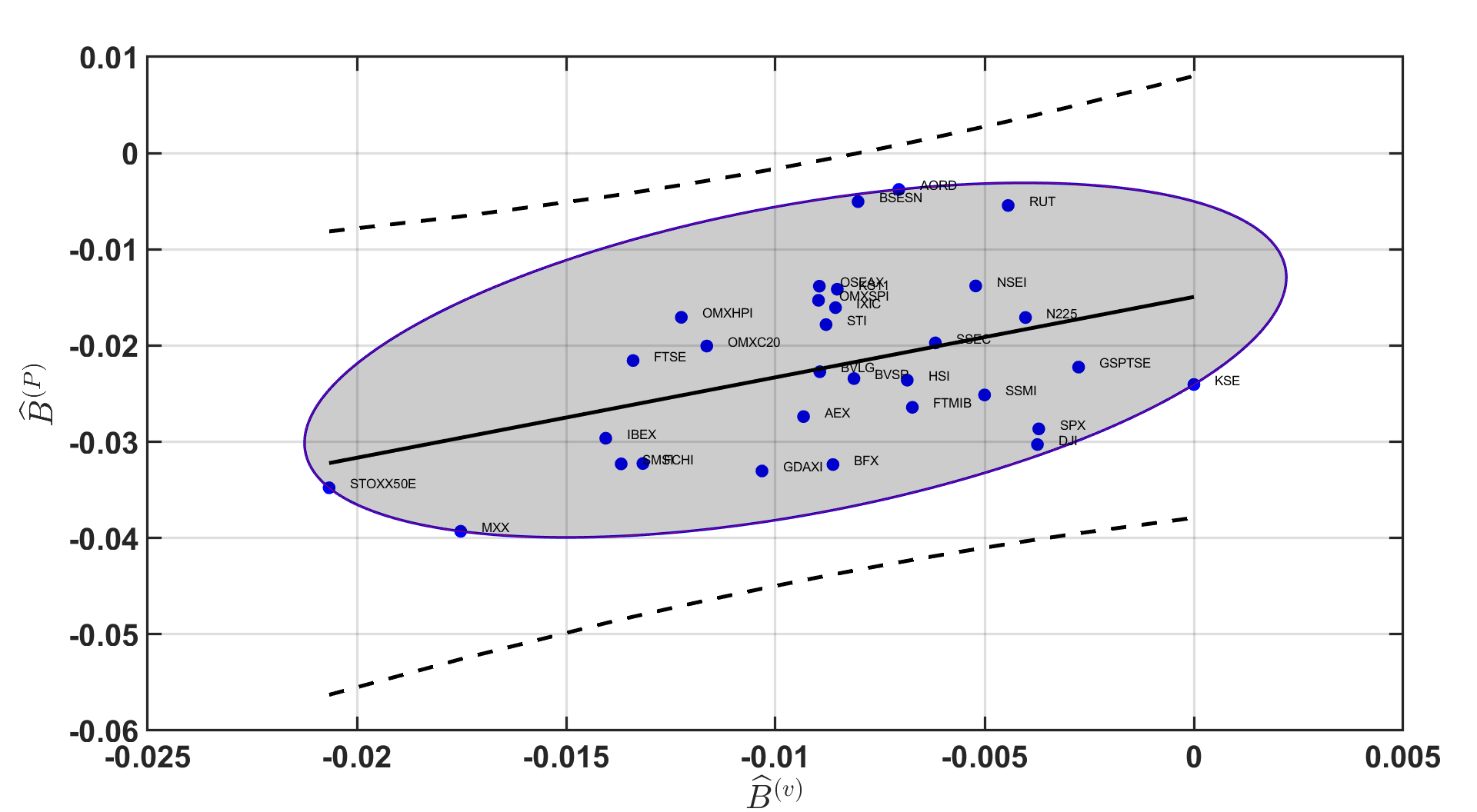} 
		\end{center}
		\caption{Estimated multiscaling proxy of the prices $\widehat{B}^{(P)}$ as function of volatility multiscaling $\widehat{B}^{(v)}$ (realized variance at 5 min frequency). Pearson correlation coefficient $\rho$ is $0.41$ and Spearman correlation coefficient $\rho_S$ is $0.33$. No outlier has been detected. Description as in caption of Figure \ref{fig1}.}
\label{fig_apx13}
\end{figure}

	\begin{figure}[H]
		\begin{center}	
			
			\includegraphics[width=1\textwidth,height=0.35\textheight]{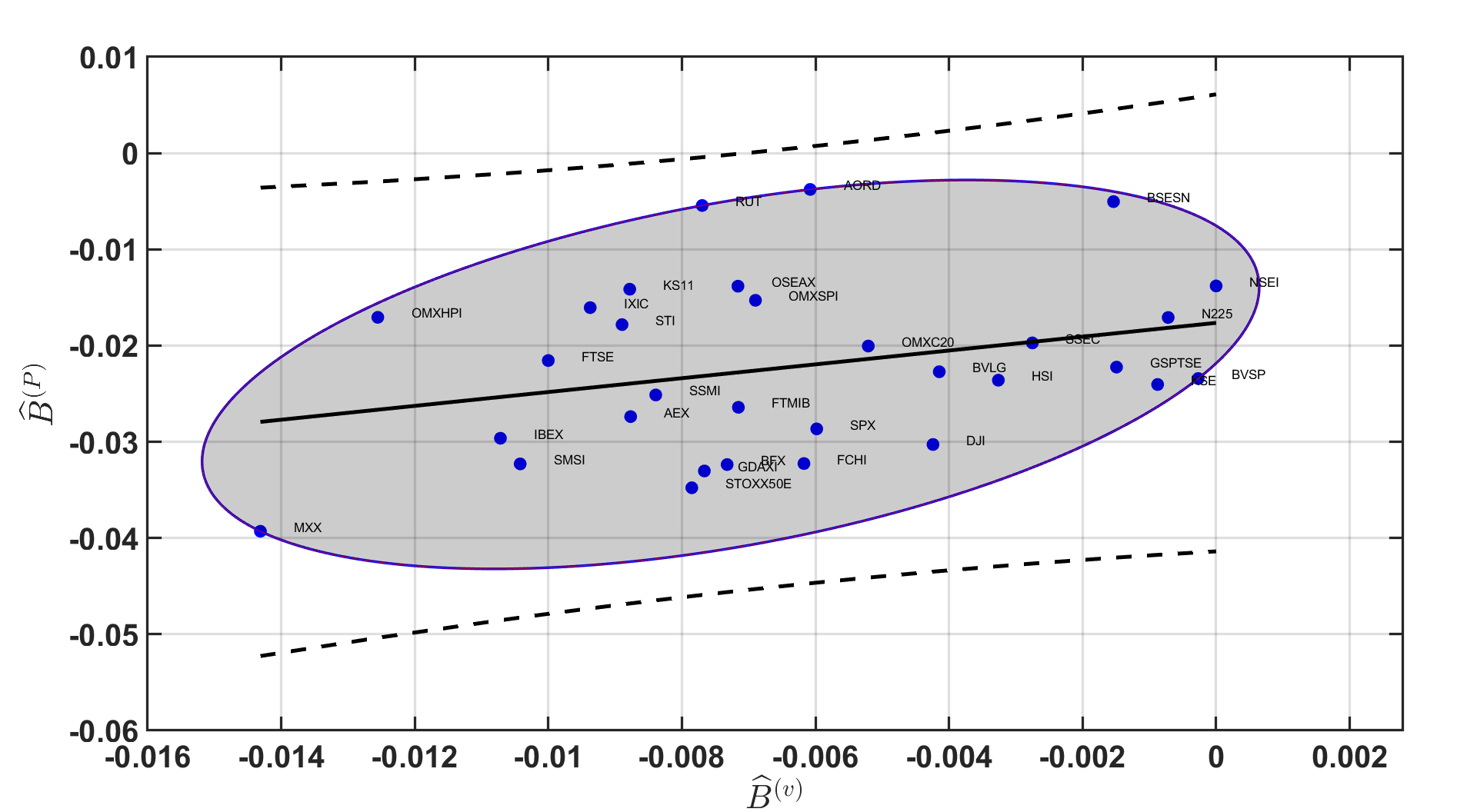} 
		\end{center}
		\caption{Estimated multiscaling proxy of the prices $\widehat{B}^{(P)}$ as function of volatility multiscaling $\widehat{B}^{(v)}$ (realized semi-variance at 5 min frequency). Pearson correlation coefficient $\rho$ is $0.30$ and Spearman correlation coefficient $\rho_S$ is $0.25$. No outlier has been detected. Description as in caption of Figure \ref{fig1}.}
\label{fig_apx14}
\end{figure}

	\begin{figure}[H]
		\begin{center}	
			
			\includegraphics[width=1\textwidth,height=0.35\textheight]{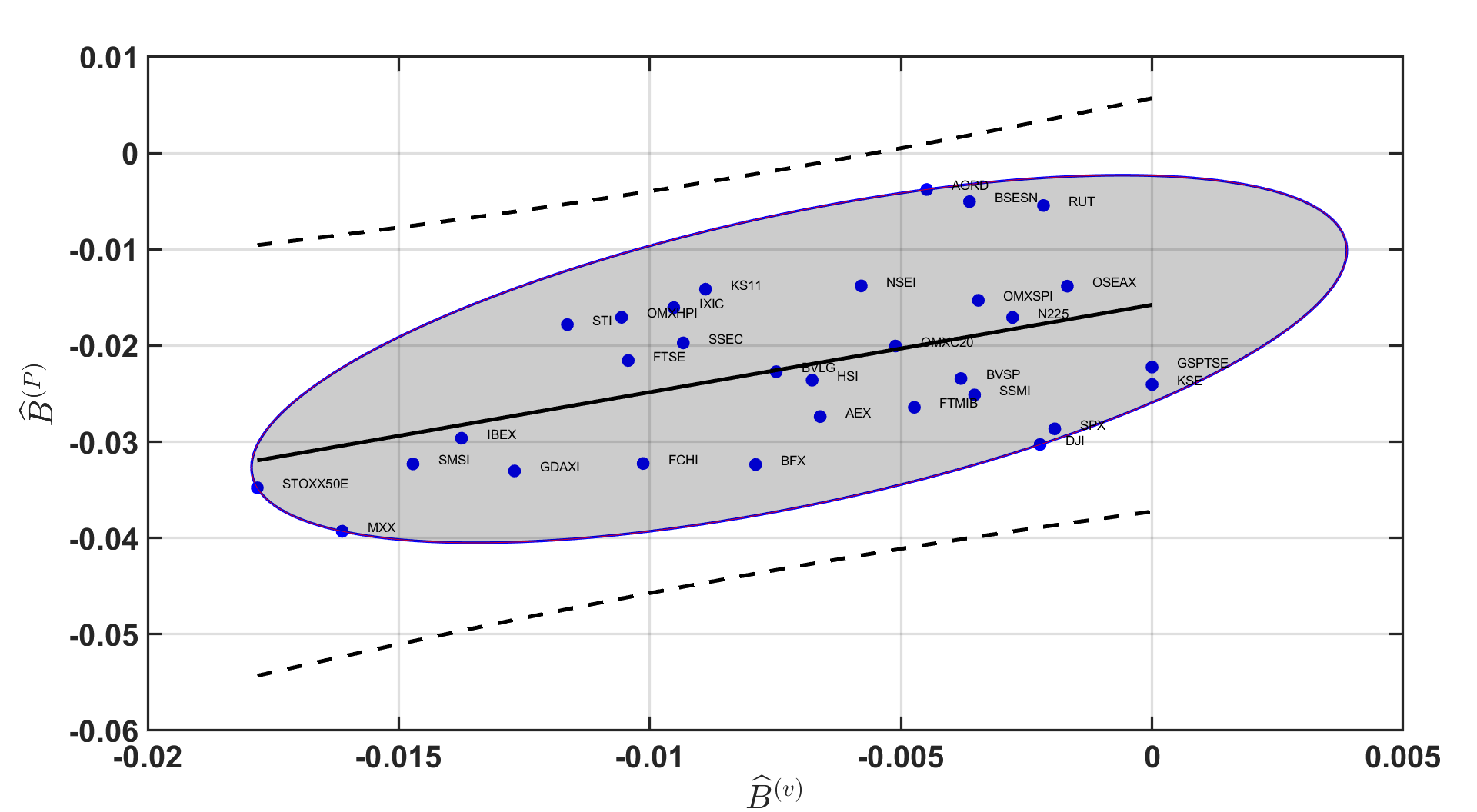} 
		\end{center}
		\caption{Estimated multiscaling proxy of the prices $\widehat{B}^{(P)}$ as function of volatility multiscaling $\widehat{B}^{(v)}$ (bipower variation at 5 min frequency). Pearson correlation coefficient $\rho$ is $0.49$ and Spearman correlation coefficient $\rho_S$ is $0.40$. No outlier has been detected. Description as in caption of Figure \ref{fig1}.}
\label{fig_apx15}
\end{figure}

Figures \ref{fig_apx13}-\ref{fig_apx15} show that the correlation between the multiscaling features of the volatility and prices processes is positive. Furthermore, the Pearson correlation is statistically significant across different volatility measures, while the Spearman correlation is not statistically significant for all the volatility measures. 

\subsection{Simulated data with empirical Hurst exponent}\label{add_res_real_sim}

	\begin{figure}[H]
		\begin{center}	
			
			\includegraphics[width=1\textwidth,height=0.44\textheight]{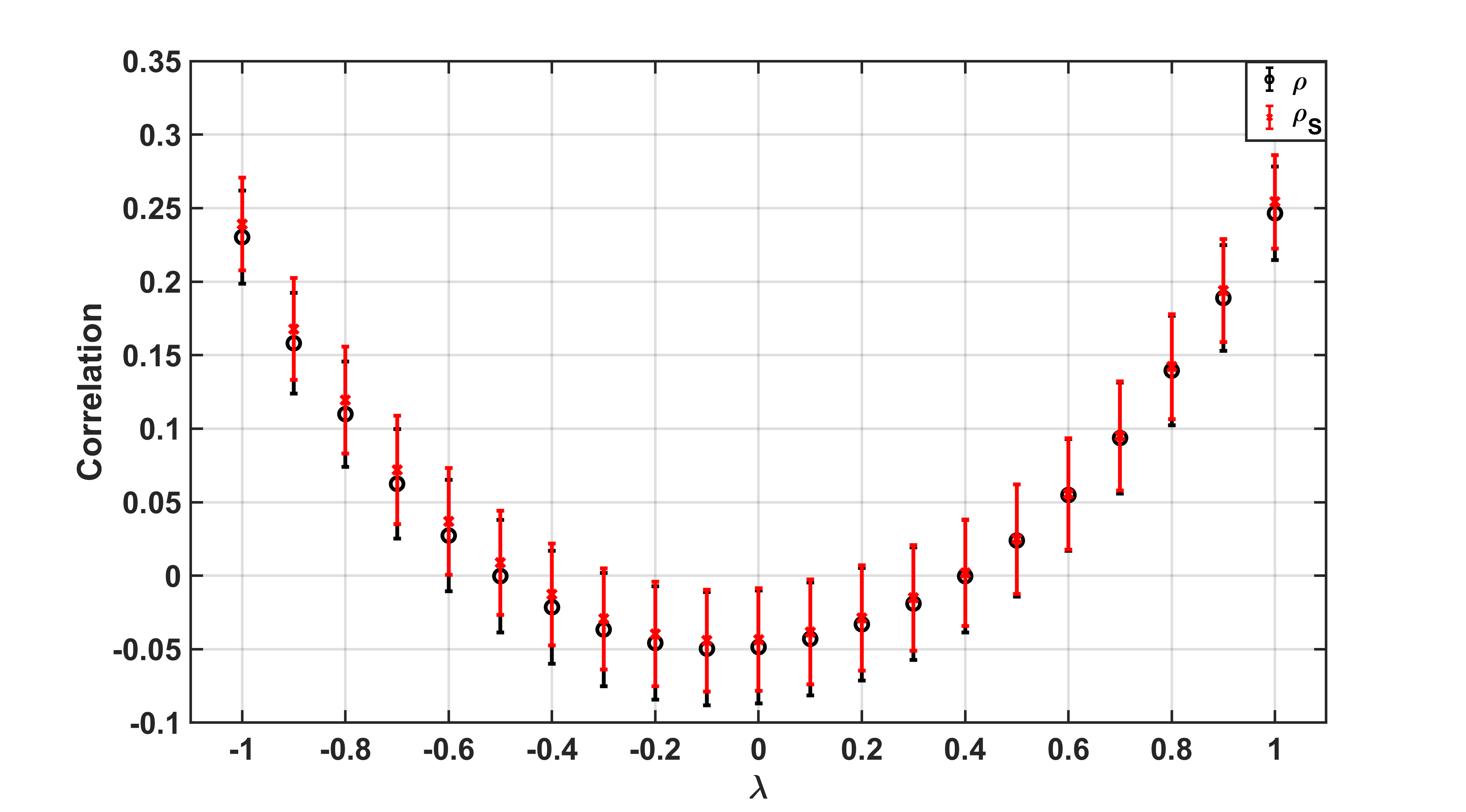} 
		\end{center}
		\caption{Correlation coefficients between $\widehat{H}^{(v)}$ and $\widehat{H}^{(P)}$ as function of $\lambda$ for the rBergomi model with $H$ taken from Table \ref{res_scaling}, $\xi=0.01$, $\eta=1.9$. Description as reported in caption of Figure \ref{fig1b}.}
		\label{fig_realH_HH}
	\end{figure}

	\begin{figure}[H]
		\begin{center}	
			
			\includegraphics[width=1\textwidth,height=0.44\textheight]{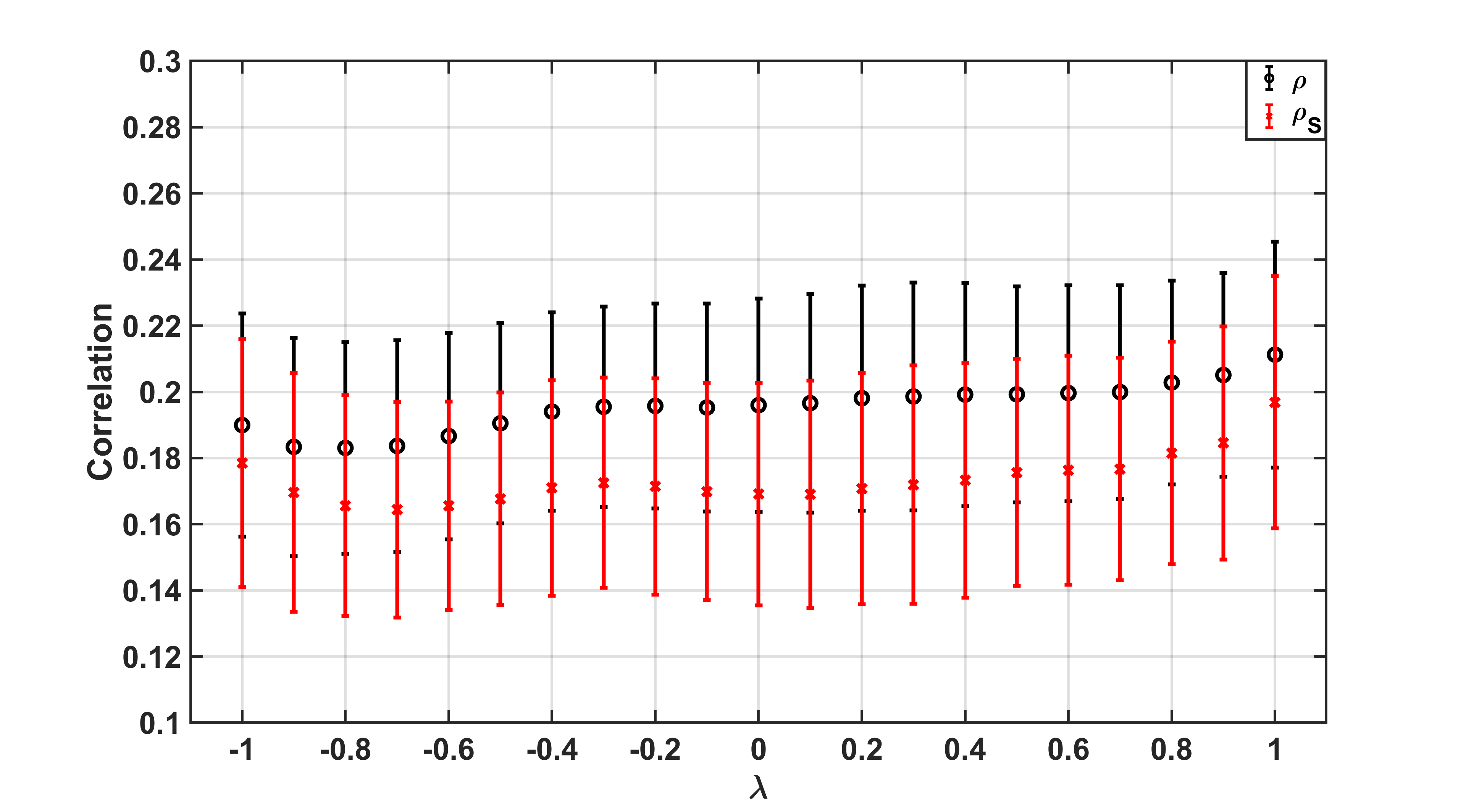} 
		\end{center}
		\caption{Correlation coefficients between $\widehat{B}^{(v)}$ and $\widehat{B}^{(P)}$ as function of $\lambda$ for the rBergomi model with $H$ taken from Table \ref{res_scaling}, $\xi=0.01$, $\eta=1.9$. Description as reported in the caption of Figure \ref{fig1b}.}
		\label{fig_realH_BB}
	\end{figure}

Figure \ref{fig_realH_HH} shows that the rBergomi produces a correlation between the Hurst exponents of the volatility and the Hurst exponent of the prices process which is in line with the input correlation $\lambda$ of the Brownian motions. Indeed, the correlation between $\widehat{H}^{(v)}$ and $\widehat{H}^{(P)}$ is not statistically significant at $5\%$ for values of $\lambda$ between $-0.6$ and $0.5$. On the other hand, the correlation between the multiscaling features is stable around $\sim0.18$ irrespective of $\lambda$, confirming that $\lambda$ does not have a direct effect on the multiscaling properties of volatility and prices processes.

\newpage

\section{Outlier identification procedure}\label{outliers_sec}
\setcounter{table}{0}
\setcounter{figure}{0}

 Let's $X \in \mathcal{R}^{N \times 2}$ be the bivariate dataset (in our case $X$ is composed by $\widehat{H}^{(v)}$ and $\widehat{B}^{(P)}$) composed by $N$ datapoints with indices $I\in \{1,\dots,N\}$. The procedure is as follow:

\begin{enumerate}
    \item Compute the Minimum Covariance Determinant (MCD) of the dataset \citep{hubert2018minimum}
    \item Compute $\mu$ as the center of the data scatter cloud given by the MCD \citep{pernet2013robust}
    \item Compute the (Euclidean) distance $D_i$ to the center of the data, i.e. $X-\mu$ for all set of points $i=1,\dots,N$
    \item Use the (corrected) Boxplot rule by \citep{carling2000resistant} to detect the outliers in $D_i$
    \item Define the set of outliers as $o$ and the set of datapoints without outliers as $l=I\setminus o$
    \item Compute the robust correlation coefficient $\tilde{\rho}=\rho(X_l)$, where $X_l$ is the set of bivariate datapoints filtered by outliers. 
\end{enumerate}

For the parameter choice in the various steps of the procedure, we use the optimal ones described in \citep{pernet2013robust,hubert2018minimum,carling2000resistant}. It is important to notice that being the MCD is a robust method to compute a scatter matrix (covariance matrix), a robust correlation coefficient can be computed directly from it \citep{hubert2018minimum}.\\
Regarding the procedure used to detect outliers, it is possible that some bivariate datapoints are outliers for one specification (volatility measure and but not for another. In order to remove only the very strong outlier(s) which affect all the specifications, we define as the set of outliers the intersection between the outliers found across different volatility measures.\footnote{A less stringent rule would be to classify as outliers the ones which result to be an outlier for the majority of the specifications rather than for all.} Define $m$ as the index of a specific volatility measure. We define $o_{m}$ as the set of outliers for a specific volatility measure $m$. The overall outlier set is computed as:

\begin{equation}
\tilde{o}=\bigcap_m o_{m},
\end{equation}

where $\tilde{o}$ is the set of outliers for all the specifications.

\end{document}